\documentclass[12pt,preprint]{aastex}

\def\msol{\hbox{\kern 0.20em $M_\odot$}}
\def\lsol{\hbox{\kern 0.20em $L_\odot$}}
\def\rsol{\hbox{\kern 0.20em $R_\odot$}}
\def\sr{\hbox{\kern 0.20em sr}}
\def\srmu{\hbox{\kern 0.20em sr$^{-1}$}}
 
\def\g{\hbox{\kern 0.20em g}}
\def\gmu{\hbox{\kern 0.20em g$^{-1}$}}
\def\kg{\hbox{\kern 0.20em kg}}
\def\pc{\hbox{\kern 0.20em pc}}
 
\def\mum{\hbox{\kern 0.20em $\mu$m}}
\def\mumd{\hbox{\kern 0.20em $\mu$m$^{-2}$}}
\def\cm{\hbox{\kern 0.20em cm}}
\def\m{\hbox{\kern 0.20em m}}
\def\km{\hbox{\kern 0.20em km}}
\def\nm{\hbox{\kern 0.20em nm}}
 
\def\s{\hbox{\kern 0.20em s}}
\def\h{\hbox{\kern 0.20em h}}
\def\sec{\hbox{\kern 0.20em sec}}
\def\min{\hbox {\kern 0.20em min}}
\def\smu{\hbox{\kern 0.20em s$^{-1}$}}
\def\smd{\hbox{\kern 0.20em s$^{-2}$}}
\def\an{\hbox{\kern 0.20em an}}
\def\anmu{\hbox{\kern 0.20em an$^{-1}$}}
\def\deg{\hbox{\kern 0.20em $^{\rm o}$}}
\def\yr{\hbox{\kern 0.20em yr}}
\def\yrmu{\hbox{\kern 0.20em yr$^{-1}$}}
\def\Myr{\hbox{\kern 0.20em Myr}}
\def\Mymu{\hbox{\kern 0.20em Myr$^{-1}$}}
\def\K{\hbox{\kern 0.20em K}}
\def\pcmu{\hbox{\kern 0.20em pc$^{-1}$}}
\def\pcmd{\hbox{\kern 0.20em pc$^{-2}$}}
\def\pcmt{\hbox{\kern 0.20em pc$^{-3}$}}
\def\kms{\hbox{\kern 0.20em km\kern 0.20em s$^{-1}$}}
\def\kmpd{\hbox{\kern 0.20em km$^{2}$}}
\def\kpc{\hbox{\kern 0.20em kpc}}
\def\cms{\hbox{\kern 0.20em cm\kern 0.20em s$^{-1}$}}
\def\erg{\hbox{\kern 0.20em erg}}
\def\ergs{\hbox{\kern 0.20em erg}}
\def\cmpd{\hbox{\kern 0.20em cm$^2$}}
\def\cmmd{\hbox{\kern 0.20em cm$^{-2}$}}
\def\cmms{\hbox{\kern 0.20em cm$^{-6}$}}
\def\cmpt{\hbox{\kern 0.20em cm$^3$}}
\def\cmmt{\hbox{\kern 0.20em cm$^{-3}$}}
\def\mpd{\hbox{\kern 0.20em m$^2$}}
\def\mmd{\hbox{\kern 0.20em m$^{-2}$}}
\def\mpt{\hbox{\kern 0.20em m$^3$}}
\def\mmt{\hbox{\kern 0.20em m$^{-3}$}}
\def\mujy{\hbox{\kern 0.20em $\mu$Jy}}
\def\mjy{\hbox{\kern 0.20em mJy}}
\def\Mj{\hbox{\kern 0.20em MJy}}
\def\jy{\hbox{\kern 0.20em Jy}}
\def\ghz{\hbox{\kern 0.20em GHz}}
\def\srmd{\hbox{\kern 0.20em sr$^{-1}$}}
\def \kms{km~$\rm{s}^{-1}$}

\def \mum{$\mu$m}

\def\G{\hbox{\kern 0.20em G}}

\def\h13cop{\hbox{H$^{13}$CO$^{+}$}}

\def\S+{\hbox{S{\small II}}}

\shorttitle{WIYN/Hydra Spectroscopic Survey in the Spitzer FLS}
\shortauthors{Marleau et al.}

\lefthead{Marleau et al.}

\righthead{WIYN/Hydra Spectroscopic Survey in the Spitzer FLS}

\begin{document}

\newcommand{\jfourteen}{\hbox{$J=14\rightarrow 13$}} 
\title{Spectroscopic Survey of 1.4~GHz and 24 Micron Sources
in the Spitzer First Look Survey with WIYN/Hydra\altaffilmark{1}}

\author{F.R.\ Marleau\altaffilmark{2}, D.\ Fadda\altaffilmark{3},
P.N.\ Appleton\altaffilmark{3}, A.\ Noriega-Crespo\altaffilmark{2},
M.\ Im\altaffilmark{4}, and D.\ Clancy\altaffilmark{2}}

\altaffiltext{1}{based on observations obtained at the WIYN telescope
at the Kitt Peak National Observatory, National Optical Astronomy
Observatory, which is operated by the Association of Universities
for Research in Astronomy, Inc., under cooperative agreement
with the National Science Foundation.}

\altaffiltext{2}{SPITZER Science Center, California Institute of
Technology, CA 91125 USA.}

\altaffiltext{3}{NASA Herschel Science Center, California Institute 
of Technology, CA 91125 USA.}

\altaffiltext{4}{Department of Physics \& Astronomy, FPRD, Seoul
National University, Seoul, Korea.}

\begin{abstract}
We present an optical spectroscopic survey of 24~\mum\ and 1.4~GHz
sources, detected in the Spitzer Extragalactic First Look Survey
(FLS), using the multi-fiber spectrograph, Hydra, on the WIYN
telescope. We have obtained spectra for 772 sources, with flux
densities above 0.15~mJy in the infrared, and 0.09~mJy in the
radio. The redshifts measured in this survey are mostly in the range
$0 < z < 0.4$, with a distribution peaking at $z \sim 0.2$. Detailed
spectral analysis of our sources reveals that the majority are
emission-line star-forming galaxies, with star formation rates in the
range 0.2-200 M$_\odot$ yr$^{-1}$. The rates estimated from the
H$\alpha$ line fluxes are found to be on average consistent with
those derived from the 1.4~GHz luminosities. For these star-forming
systems, we find that the 24~\mum\ and 1.4~GHz flux densities follow
an infrared-radio correlation, that can be characterized by a value of
q$_{24}$ = 0.83, with a 1-sigma scatter of 0.31. Our WIYN/Hydra
database of spectra complements nicely those obtained by the Sloan
Digital Sky Survey, in the region at lower redshift, as well as the
MMT/Hectospec survey of Papovich et al.\ (2006), and brings the
redshift completeness to 70\% for sources brighter than 2~mJy at
24~\mum. Applying the classical 1/V$_{max}$ method, we derive new
24~\mum\ and 1.4~GHz luminosity functions, using all known redshifts
in the FLS. We find evidence for evolution in both the 1.4~GHz and
24~\mum\ luminosity functions in the redshift range $0 < z < 1$. The
redshift catalog and spectra presented in this paper are available at
the {\em Spitzer} FLS website.
\end{abstract}


\section{Introduction}

With the advent of space infrared missions in the last two decades,
the understanding of galaxy formation and evolution has undergone a
dramatic advance. Prior to this, the state of knowledge concerning the
subject had been critically limited to the information which could be
obtained mainly from the ultraviolet and optical (e.g., Lilly et al.\
1995, 1996; Madau et al.\ 1996). While such studies allowed for some
significant advances in the understanding of these processes, the
sparsity of detailed information from the infrared presented a major
obstacle to any significant further progress.

The relative importance of the infrared, with regard to galaxy
evolution, derives primarily from the fact that the majority of light
emitted by nascent stars, in the characteristically dusty star-forming
regions of galaxies, is reprocessed by the surrounding dust and
re-emitted at infrared wavelengths (see, e.g., Calzetti, Kinney \&
Storchi-Bergmann 1994). Therefore, a great deal of the information
concerning star formation, its history, and hence galaxy evolution, is
expected to be encoded in the infrared. In the local Universe, such
infrared emission is known to constitute about one third of the total
light emitted from galaxies (Soifer \& Neugebauer 1991). However, as
can be inferred from the cosmic background light, this contribution
increases significantly with redshift, accounting for half of the
total energy produced by extragalactic sources, and is due to a
corresponding increase in dust-obscured star formation and/or
accretion activity (Dole et al.\ 2006; Puget et al.\ 1996; see also
Hauser \& Dwek 2001 for a review).

With the launch of the {\em Infrared Astronomical Satellite} (IRAS) in
January 1983, a number of studies of faint sources provided the first
evidence for evolution in the infrared galaxy population (Hacking,
Condon \& Houck 1987; Lonsdale et al.\ 1990). However, it wasn't until
the launch of the {\em Infrared Space Observatory} (ISO) that the
first detailed studies of the deep Universe were undertaken, providing
new insights on galaxy evolution based on higher redshift sources. In
particular, discoveries such as the excess of faint sources with
respect to no-evolution models, measured in the 15~\mum\ number counts
(e.g., Elbaz et al.\ 1999; Flores et al.\ 1999; Lari et al.\ 2001;
Metcalfe et al.\ 2003), led to several evolutionary models for
infrared galaxies (see, e.g., Franceschini et al.\ 2001; Chary \&
Elbaz 2001; Lagache et al.\ 2003). In addition, an initial attempt was
also made with ISO to study the 15~\mum\ local luminosity function
(Pozzi et al.\ 2004).  However, the success of this study was somewhat
hampered by the insufficient quality of the data for the task, and by
limited redshift information.

More recently, the {\em Spitzer Space Telescope}\footnote{The {\em
Spitzer Space Telescope} is operated by the Jet Propulsion Laboratory
(JPL), California Institute of Technology, under NASA contract 1407.}
opened the door to further advances and discoveries, with its
significantly improved detector technology, providing both better
resolution and sensitivity (Werner et al.\ 2004). The first survey to
be performed by {\em Spitzer}, the public First Look Survey (FLS),
included an extragalactic component which consisted of shallow
observations of a 4.4 sq.\ deg.\ field, centered at $\alpha(2000)$ =
17$^h$18$^m$00$^s$, $\delta(2000)$ = 59\arcdeg 30\arcmin
00\arcsec. The observations were obtained using the four Infrared
Array Camera (IRAC) channels (Fazio et al.\ 2004), and the three
Multiband Imaging Photometer (MIPS) bands (Rieke et al.\ 2004). An
important aspect of the FLS was that it was also augmented by a large
pool of ancillary data.  Specifically, the survey was complemented at
other wavelengths by Sloan Digital Sky Survey (SDSS) and Kitt Peak
National Observatory (KPNO) observations, obtained in several optical
bands (Hogg et al.\ 2007, in preparation; Fadda et al.\ 2004); by
deeper optical and near-infrared imaging, obtained with the
Canada-France-Hawaii Telescope (CFHT) and the 200 inch telescope at
Palomar (Shim et al.\ 2006; Glassman et al.\ 2007, in preparation); by
imaging of the same field with both the Hubble Space Telescope (HST)
and the Galaxy Evolution Explorer (GALEX) (see, e.g., Bridge et al.\
2007; Damjanov et al.\ 2006); and finally, by 1.4~GHz radio data,
obtained with the Very Large Array (VLA), contributing the deepest
radio survey in such a wide area to date (Condon et al.\ 2003).

Some of the new science made possible by this survey included the
estimation of the 24~\mum\ galaxy number counts (Marleau et al.\ 2004)
and the MMT/Hectospec redshift survey of Papovich et al.\
(2006). Galaxy evolution was inferred from the 24~\mum\ counts
measured in the FLS and other deep {\em Spitzer} surveys (Marleau et
al.\ 2004; Papovich et al.\ 2004), and this led to a revision of
existing models (e.g., Lagache et al.\ 2004). Following the FLS, other {\em
Spitzer} surveys were conducted to study the infrared galaxy
population. For example, the 24~\mum\ luminosity function was
constructed in the {\em Spitzer} Wide-area Infrared Extragalactic
(SWIRE) survey (Babbedge et al.\ 2006), and the Chandra Deep Field
South (CDFS) and Hubble Deep Field North (HDF-N) surveys
(P\'erez-Gonz\'alez et al.\ 2005). However, an important limitation of
these studies was the fact that they were based completely on
photometric redshift determination, and so lacked the reliability of
spectroscopic surveys. Indeed, although multi-wavelength imaging
provides a wealth of information regarding the nature of extragalactic
sources, spectroscopic observations are in the end necessary to probe
deeper into the nature of these sources, by providing a more accurate
and detailed determination of the chemical properties, redshift
distributions, luminosity functions, star-formation rates and
evolutionary effects. The survey presented in this paper, which we
shall now detail, aimed at doing just that.

Our program of observations and research commenced in 2002, prior to
the August 2003 launch of {\em Spitzer}, with follow-up optical
spectroscopy, using the Hydra spectrograph on the WIYN
telescope\footnote{The WIYN Observatory is a joint facility of the
University of Wisconsin-Madison, Indiana University, Yale University,
and the National Optical Astronomy Observatories.} at KPNO, and
continued through until 2005. The first sources to be targeted by our
survey were radio sources detected with the VLA at 1.4~GHz. Based on
the infrared-radio correlation (see, e.g., Dickey \& Salpeter 1984; de
Jong et al.\ 1985; Helou et al.\ 1985; Condon \& Broderick 1986), we
expected these sources to be infrared-bright sources and, therefore,
to be detected by {\em Spitzer}.

A study of the galaxy radio population offers the additional advantage
that radio emission does not suffer from extinction and, therefore, is
a more accurate tracer of both star formation and active galactic
nuclei (AGN) activity. The 1.4~GHz luminosity function of star-forming
galaxies and AGNs is well known only for the relatively local Universe
(e.g., Condon, Cotton \& Broderick 2002; Sadler et al.\ 2002). In
fact, very little is known about the evolution of the radio luminosity
function as, until recently, the deepest large scale radio survey
reached 2.5~mJy. Moreover, attempts at measuring evolution of the
radio luminosity function have remained inconclusive due to large
uncertainties (Machalski \& Godlowski 2000). After the FLS
observations were completed in December 2003, we targeted the infrared
sources detected at 24~\mum\ with MIPS. The database of optical
spectra we present in this paper provides a firm basis for the study
of the bright end of the 1.4~GHz and 24~\mum\ luminosity function, and
can also be used, for example, as the training set for photometric
redshifts estimation, using the neural network method (see, e.g.,
Collister \& Lahav 2004).

The paper is organized as follows: WIYN observations and data
reduction are presented in Section~2, spectral analysis and
completeness of the survey are discussed in Section~3, measurement of
emission lines, diagnostic diagrams and star formation rates are
presented in Section~4, luminosities and the infrared-radio
correlation are computed in Section~5, and in Section~6, luminosity
functions at 1.4~GHz and 24~\mum\ are derived. Section~7 describes how
to access the database of the WIYN/Hydra spectra. Finally, Section~8
summarizes the main results of the paper.

Throughout this paper we assume $H_0 = 75$ km s$^{-1}$ Mpc$^{-1}$,
$\Omega_M = 0.3$, and $\Omega_{\Lambda} = 0.7$.

\section{Observations and Data Reduction}

\subsection{WIYN/Hydra Target Selection}

\begin{figure}
\centerline{
\includegraphics[width=300pt,height=300pt,angle=0]{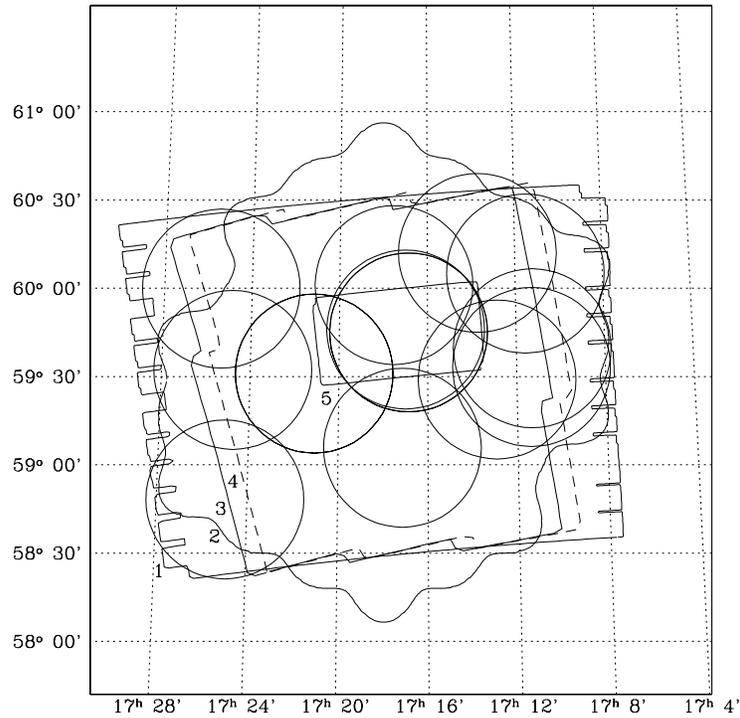}
}
\caption{\label{fig:fields} Coverage of 1) infrared {\em Spitzer}
MIPS-24~\mum\ main FLS field (Fadda et al.\ 2006), 2) VLA 1.4~GHz
(Condon et al.\ 2003), 3) infrared {\em Spitzer} IRAC channels 1 and
3 (Lacy et al.\ 2005), 4) infrared {\em Spitzer} IRAC channels 2 and
4 (Lacy et al.\ 2005), and 5) infrared {\em Spitzer} MIPS-24~\mum\ deeper
FLS verification region (Fadda et al.\ 2006). The {\em solid circles}
outline the 15 fields (see Table~\ref{tbl:fields}) where our
WIYN/Hydra spectroscopic targets were selected.}
\end{figure}

Our WIYN/Hydra target sample was selected from two datasets of sources
detected in the FLS. The first set was comprised of sources detected
at 1.4~GHz and with an R-band magnitude $<23$. This set was used
during our 2002 and 2003 observing runs, prior to the {\em Spitzer}
launch (2003 August 25). The second set was constructed from the
24~\mum\ sources, using the same R-band magnitude cut. These were
targeted during our 2005 observing run after the {\em Spitzer}
observations of the FLS were completed (2003 December 9-11). A few
companions of 1.4~GHz and 24~\mum\ sources were also part of the
sample. Our list of targets was divided into 15 separate pointings
(see Table~\ref{tbl:fields}). Each Hydra field covered a circular area
0.9 degrees in diameter, as shown in Figure~\ref{fig:fields}. Note
that some of the fields seen in Figure~\ref{fig:fields} are
overlapping. In these fields, the source density was larger than the
maximum number of allowed fibers and therefore these regions were
observed more than once.

\subsection{WIYN/Hydra Observational Setup}

Optical spectra of the selected radio and infrared galaxies were
obtained with the Hydra Multi-Object Spectrograph on the WIYN 3.5m
telescope at the Kitt Peak National Observatory\footnote{Kitt Peak and
the National Optical Astronomy Observatory are operated by the
Association of Universities for Research in Astronomy, Inc., under
cooperative agreement with the National Science Foundation.}. Hydra
has a total of 288 fiber positions (numbered $0-287$), distributed
evenly around the focal plate with every third position being a red
fiber. Of these $\sim$ 98 red fibers, only 93 were working properly at
the time of our observations.

Each red fiber was assigned a target by taking into account a minimum
fiber-to-fiber placement distance of 37 arcsec. The fibers in each
field were assigned to either a target, the sky, or a guide star. On
average, ten fibers were assigned to the sky in each pointing. An
additional set of eleven fibers were available for guide stars. Target
selection was carried out by running the assignment code {\em
whydra}. The code optimized fiber placement based on a set of
user-defined weights, subject to a set of fiber placement rules. Once
at the telescope, the instrument program {\em hydrawiyn} made the
final assignment of fibers to targets. We found that typically,
$60-70$ fibers were assigned to selected survey targets in each field.

Multi-fiber spectroscopy, using 2 arcsec diameter fibers, required
that the relative errors of the target coordinates be better than 0.5
arcsec. To achieve this precision, we used coordinates from the SDSS
and the KPNO R-band images (Fadda et al.\ 2004). The relative errors
in these optical catalogs were less than 0.1 arcsec. Also important
was the choice of guide stars, as at least three stars were selected
for each configuration. We selected guide stars with magnitudes $V =
10 - 15$ from the Tycho2 catalog (H$\o$g et al.\ 2000). The
coordinates had an absolute position error less than
60~milliarcseconds, allowing for accurate acquisition and guiding.

Our sample was observed on the nights of 2002 August 14-15, 2003 June
30-July 01, and 2005 July 17. We selected the $4000-9500$\AA\ band and
the red fiber set to observe emission lines from [OII] $\lambda$3727
to H$\alpha$ $\lambda$6563, for typical objects in our sample ($z \sim
0.2$). The red fiber set, combined with the Bench Spectrograph Camera,
was optimized for the $4500-10000$\AA\ wavelength range. However,
because of the presence of strong sky lines, the spectra redward of
$\sim$7500\AA\ were not usable. The size of the fibers was ideally
suited for our sample as each fiber covered a significant portion of a
target. We used a grating with 316~lines/mm and a blaze angle of 7
degrees, yielding a dispersion of 2.64\AA\ pixel$^{-1}$. The final
resolution, measured from the FWHM of comparison lines, was $\sim$
5\AA.

Exposure sets of 3$\times$20 minutes were used for all fields. For
each configuration we took dome flats and Cu-Ar comparison lamp
exposures for wavelength calibration. Between each field observation,
we took exposures of spectroscopic standard stars for flux
calibration.

\begin{figure}
\centerline{
\includegraphics[width=300pt,height=300pt,angle=0]{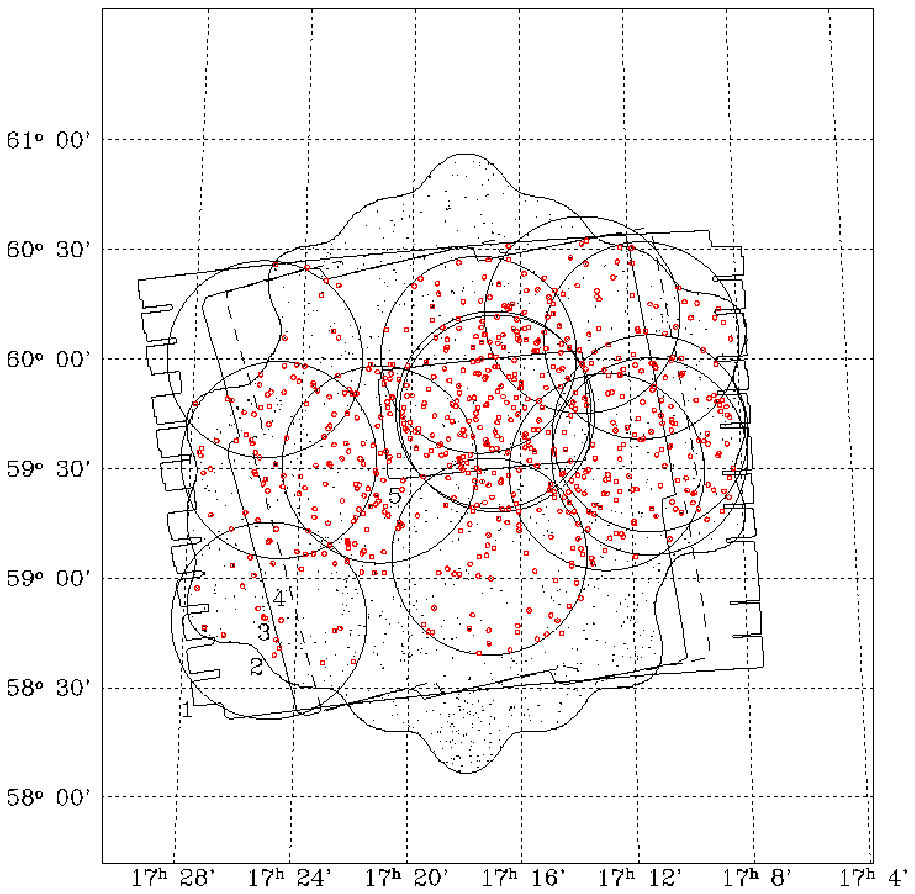} 
\includegraphics[width=300pt,height=300pt,angle=0]{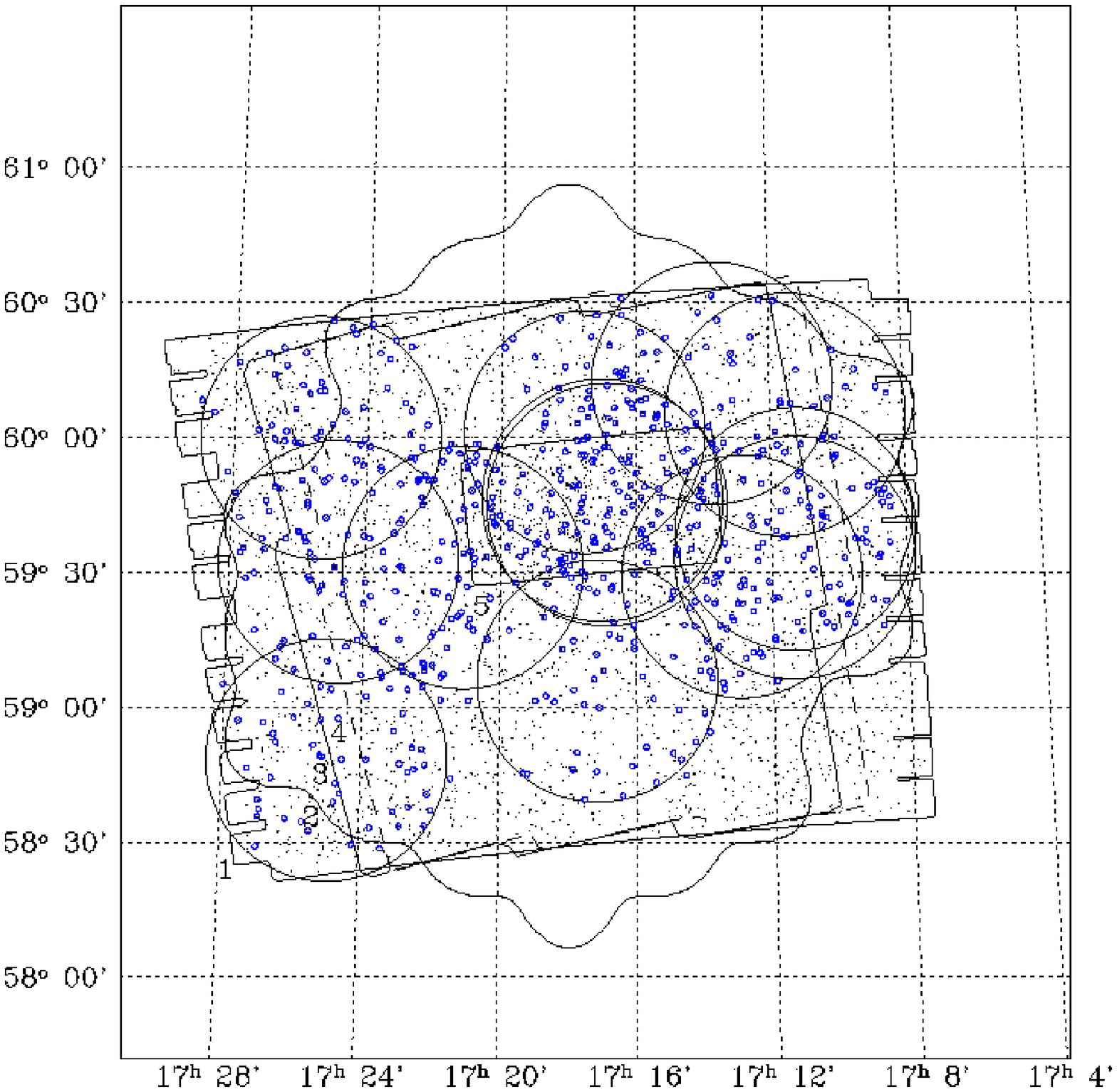}
}
\caption{\label{fig:sample} {\em Left}: WIYN/Hydra and {\em Spitzer}
FLS fields (as described in Figure~\ref{fig:fields}) and the location
of the VLA radio sources ({\em black dots}) observed with WIYN/Hydra
({\em red circles}).  {\em Right}: The location of the {\em Spitzer}
24~\mum\ sources ({\em black dots}) observed with WIYN/Hydra ({\em blue
circles}).}
\end{figure}

A total of 772 spectra were obtained. Figure~\ref{fig:sample} (left)
shows the complete radio sample from Condon et al.\ (2003) and
highlights the radio sources observed with WIYN/Hydra. The 24~\mum\
sources in the FLS catalog of Fadda et al.\ (2006) and those targeted
with WIYN/Hydra are also shown in Figure~\ref{fig:sample} (right).

\subsection{WIYN/Hydra Data Reduction}

The data reduction was carried out using a combination of tasks run
with the Interactive Data Language (IDL) and the Image Reduction and
Analysis Facility (IRAF)\footnote{IRAF is distributed by the National
Optical Astronomy Observatory, which is operated by the Association of
Universities for Research in Astronomy, Inc., under cooperative
agreement with the National Science Foundation.}. First, we created
average dome flats and comparison arc lamps for each configuration and
an average bias for each night. Images were then trimmed and the bias
was subtracted. Next, we identified the hot pixels, caused by cosmic
ray hits, by subtracting consecutive images and using the {\em
la\_cosmic} task (van Dokkum 2001). After identifying these bad
pixels, we coadded all the images relative to one configuration by
averaging their values and discarding the masked pixels.

Further processing of the science frames was done using the Hydra
reduction package {\em dohydra} (Valdes 1995), which is specifically
designed for the reduction and spectral extraction of the Hydra
multi-fiber spectra. The package allows one to extract the spectra,
compute and apply a flat-field and fiber throughput correction,
calibrate in wavelength and subtract the sky.

\begin{figure}
\centerline{
\includegraphics[width=300pt,height=300pt,angle=0]{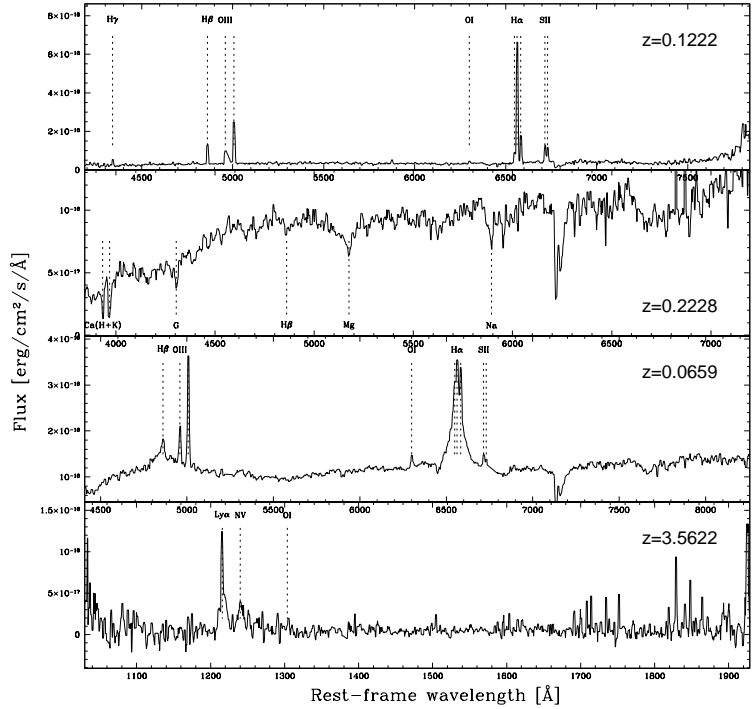}
}
\caption{\label{fig:spectra} Example spectra from our WIYN/Hydra
optical spectroscopic survey. From {\em top} to {\em bottom}: an 
emission line galaxy at $z = 0.12$, an early-type galaxy at $z = 0.22$, 
a broad emission line object at $z = 0.06$, and a quasar at $z = 3.6$.}
\end{figure}

To flux calibrate the spectra, the typical approach is to fit the
spectrum of the standard star with a high-order polynomial. However,
this removes the fine-scale features, found especially in the red part
of the spectrum, making the detection of the H$\alpha$ emission lines
of medium-redshift objects very difficult. Therefore, to calibrate in
flux, we used a different method. We filtered the spectrum of the
standard star, divided by the library spectrum, with a multiscale
transform (see, e.g., Fadda, Slezak \& Bijaoui 1998). This method
removes the high-frequency noise and conserves the significant
short-scale features. The obtained response function was then applied
to the spectra. The analysis of the spectra, including redshift and
line measurements, was carried out using an IDL code written by the
authors. Examples of the reduced spectra are shown in
Figure~\ref{fig:spectra}.

\section{Redshift Identification}

\subsection{The WIYN/Hydra Survey}

The redshifts of the infrared and radio sources in our WIYN survey
were calculated by identifying spectral features (see Section~4 below)
and taking the average. When no lines were present and the
4000\AA-break was detected, the location of the break was used to
determine the redshift. Of the 772 spectra, we measured redshifts for
498 extragalactic sources. Of these, eight were broad emission line
objects -- one of which was at $z = 3.6$ (see
Figure~\ref{fig:spectra}) -- and eleven were stars. We were able to
identify lines in 395 sources, whereas a redshift based on the
4000\AA-break was obtained for 72 sources.  The features were flagged
as ``break'', ``emission'', or ``absorption''. We visually identified
the spectroscopic classification as either ``galaxy'', ``broad
emission line'', or ``star'', and assigned each target with a redshift
quality assessment value of ``unknown'', ``tentative'' or
``good''. The results of the WIYN/Hydra spectroscopic survey are given
in an ASCII table, available in the electronic edition of the {\it
Astrophysical Journal}, with the format explained in
Table~\ref{tbl:speccat}.

Our final WIYN/Hydra redshift catalog consists of 382 radio and 412
infrared emitters, of which 322 sources are both. There are also
fifteen sources with redshifts that have neither 24~\mum\ nor 1.4~GHz
flux densities (within the flux limits of the surveys). These are
close companions of radio or infrared sources that make up a pair, and
for which we wanted to obtain both redshifts.

\subsection{Complementary Surveys}

To increase the completeness of our survey at lower redshift, we
queried the release 5 of the SDSS archive for the FLS region
(17$^h$07$^m$ $< \alpha <$ 17$^h$29$^m$, 58\arcdeg 21\arcmin $< \delta
<$ 60\arcdeg 34\arcmin). A total of 812 sources with SDSS redshifts
were retrieved. Of those, 70 sources were identified as WIYN/Hydra
sources as well, allowing us to do a redshift comparison for a
sub-sample of objects. We found excellent agreement between the
WIYN/Hydra and SDSS spectroscopic redshifts, with only two cases where
the redshifts are significantly ($|$z$_{WIYN}$-z$_{MMT}|$ $>$ 0.01)
discrepant. Given that these sources were classified as ``tentative''
under our quality index, we assigned the SDSS redshifts to these
objects.

We also retrieved the MMT/Hectospec spectroscopic catalog of Papovich
et al.\ (2006), containing 1317 measured redshifts of 24~\mum\ sources
in the FLS. In this case, we found 201 matches between the two
catalogs and agreement (as defined above) for 85\% of the matched
sources with both redshifts measured. The discrepant 15\% of the
matched sources fell either under the ``tentative'' quality assessment
or had their redshift estimated from the ``4000\AA-break'', which is
likely less reliable than identifying features. The SDSS and
MMT/Hectospec redshifts of WIYN/Hydra targets are included in
Table~\ref{tbl:speccat}.

\subsection{Optical/Radio/IR Catalogs}

The sources targeted by our survey were selected from both the 1.4~GHz
and 24~\mum\ FLS catalogs. The first sources to be targeted by our
survey, prior to the launch of {\em Spitzer}, were radio sources
detected with the VLA at 1.4~GHz. We used the catalog of Condon et
al.\ (2003) as our 1.4~GHz FLS source list, which contained 5993 radio
sources down to a flux density limit of 0.09~mJy. The 24~\mum\ target
selection was done based on the 24~\mum\ point and extended source
catalogs of Fadda et al.\ (2006). The point source catalog contained
16905 sources detected to a 5-sigma level (Table 5 of Fadda et al.\
2006, designated as ``P'' in the WIYN/Hydra redshift table). An
additional 124 extended sources were found in Table 2 of Fadda et al.\
2006 (``E'' in our table). To identify and remove stars from the point
source catalog, we queried the Sloan Digital Sky Survey, Data Release
5, for stars with r' $<$ 22. This catalog of stars was cleaned of 196
``falsely classified'' stars, by matching it to all sources with a
measured redshift in the FLS (derived from this work, the SDSS and the
work of Papovich et al.\ 2006). This allowed us to clean 933 sources
from our 24~\mum\ list, leaving a total of 16096 likely extragalactic
sources (extended sources included).

Compiling all the redshifts known in the FLS region, we constructed
the 24~\mum\ and 1.4~GHz all-surveys redshift catalogs. Of the 16096
sources at 24~\mum\, our final infrared catalog contained 412/369/1207
redshifts from WIYN/SDSS/MMT, respectively (412/305/1036 or a total of
1753 redshifts when cleaned of duplicates). Of the 5993 sources at
1.4~GHz, our final radio catalog contained 382/202/454 redshifts from
WIYN/SDSS/MMT, respectively (382/151/297 or a total of 830 redshifts
when cleaned of duplicates).

\subsection{Redshift Distribution and Completeness of the Survey}

\begin{figure}
\centerline{
\includegraphics[width=300pt,height=300pt,angle=0]{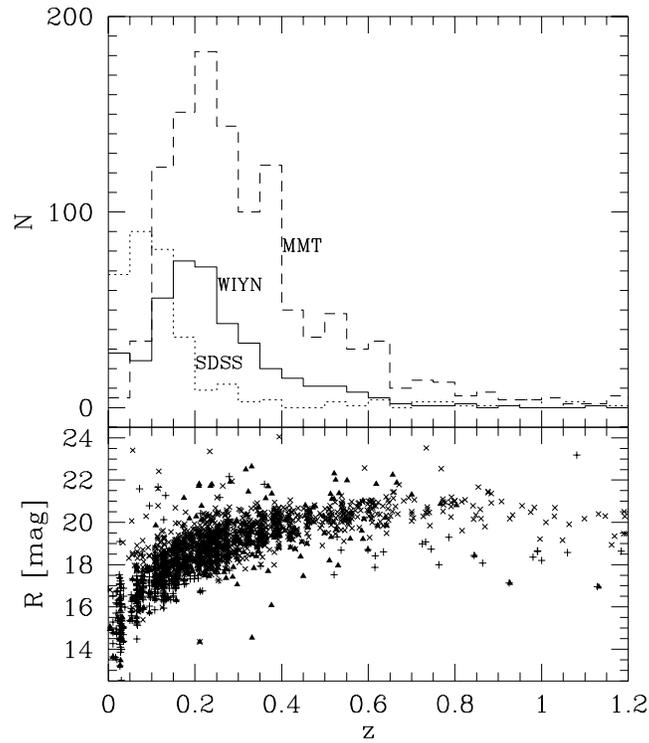}
}
\caption{\label{fig:zhist} {\em Top}: Redshift distribution of radio
and 24~\mum\ sources observed by the SDSS ({\em dotted line}), with
WIYN/Hydra ({\em solid line}) and with the MMT/Hectospec ({\em short-dash
line}). {\em Bottom}: R-band magnitude and redshift correlation for
the SDSS ({\em crosses}), WIYN/Hydra ({\em filled triangles}) and MMT
({\em diagonal crosses}) redshifts. The WIYN/Hydra survey had a
magnitude limit of R $<$ 23, corresponding roughly to a redshift limit
of z $<$ 0.8.}
\end{figure}

The redshift distribution in the FLS, including our WIYN/Hydra survey,
is shown in Figure~\ref{fig:zhist}. The majority of redshifts we
measured using the WIYN/Hydra optical spectra span the range $0 < z <
0.4$, and the distribution peaks at $z \sim 0.2$. Also displayed in
Figure~\ref{fig:zhist} are the redshift distributions of the SDSS and
the MMT/Hectospec survey. The WIYN/Hydra survey complements nicely the
SDSS at lower redshift, as well as the MMT/Hectospec survey of
Papovich et al.\ (2006), which covers a slightly larger redshift
range.

\begin{figure}
\centerline{
\includegraphics[width=220pt,height=220pt,angle=0]{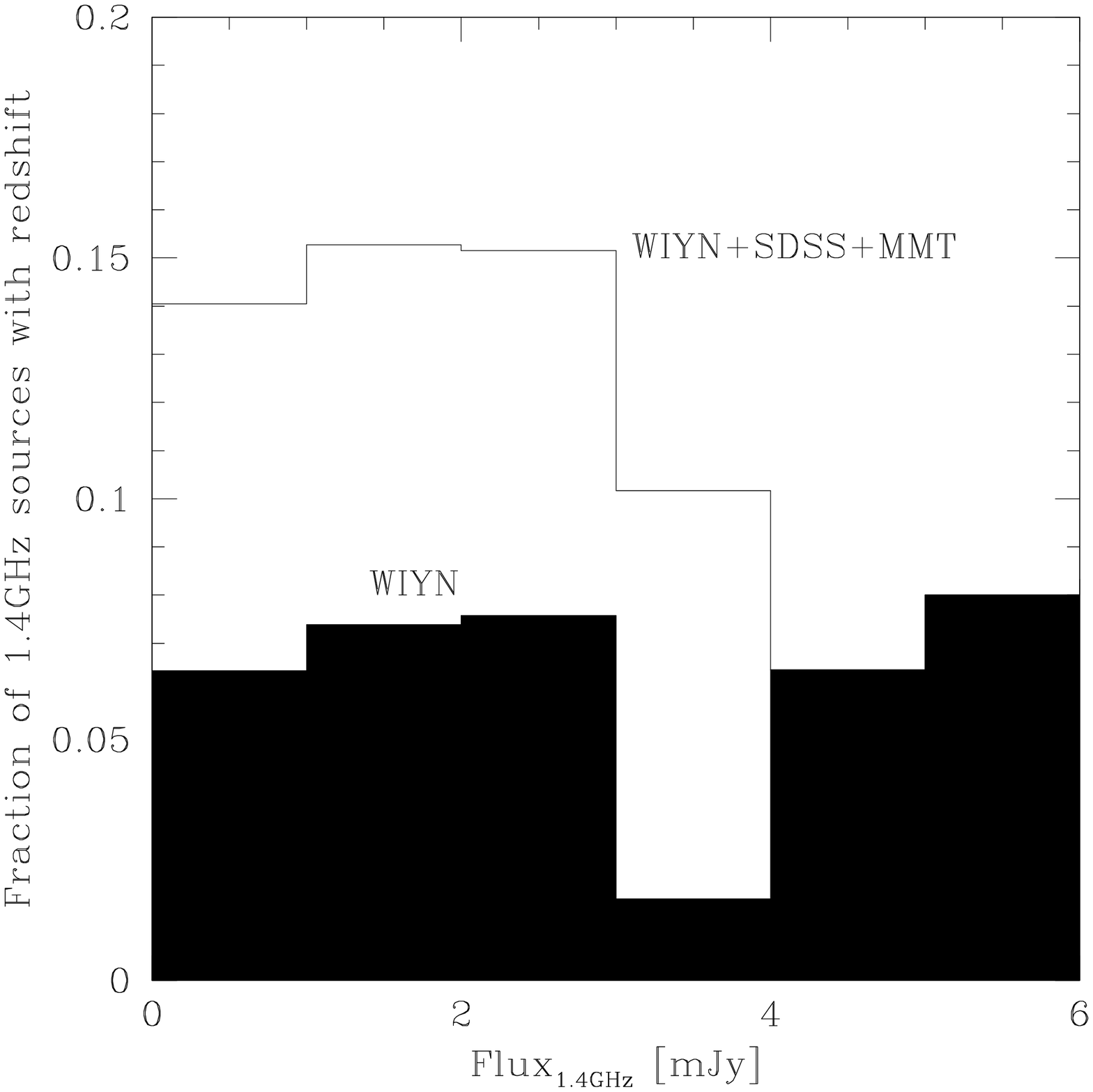}
\includegraphics[width=220pt,height=220pt,angle=0]{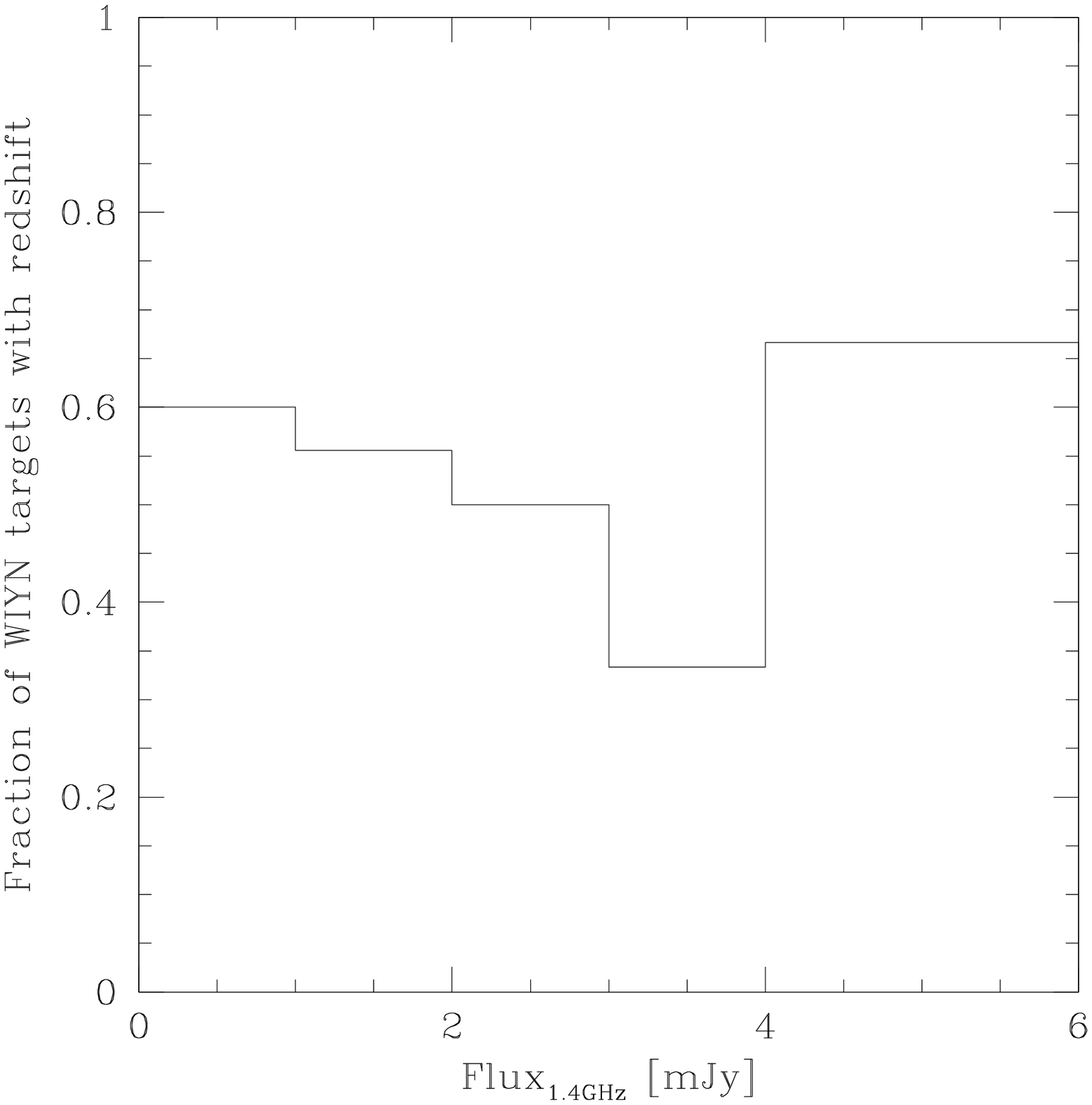}
}
\caption{\label{fig:zfracvla} {\em Left}: Fraction of 1.4~GHz sources
with spectroscopic redshifts in the FLS as a function of flux
density. {\em Right}: Fraction of WIYN/Hydra targets for which we have
successfully measured a redshift as a function of their 1.4~GHz flux
density.}
\end{figure}

The fraction of 1.4~GHz sources for which we successfully obtained a
redshift is displayed in Figure~\ref{fig:zfracvla} (left) as a
function of flux density. A total of 641 out of the 5993 sources in
this catalog were observed with WIYN/Hydra, and 382 redshifts were
successfully measured, yielding a success rate of $\sim$ 60\% (see
Figure~\ref{fig:zfracvla}, right).

\begin{figure}
\centerline{
\includegraphics[width=220pt,height=220pt,angle=0]{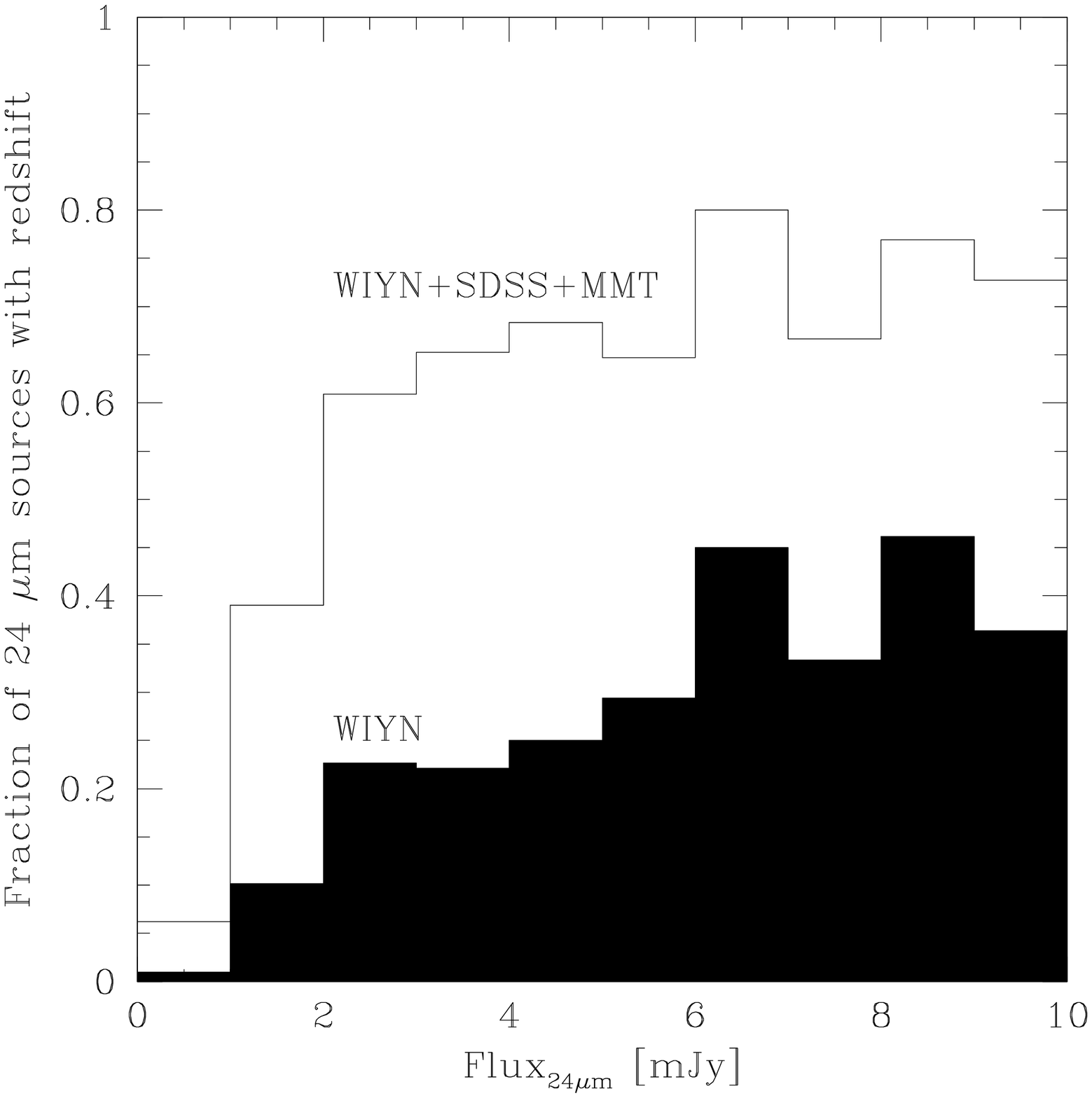}
\includegraphics[width=220pt,height=220pt,angle=0]{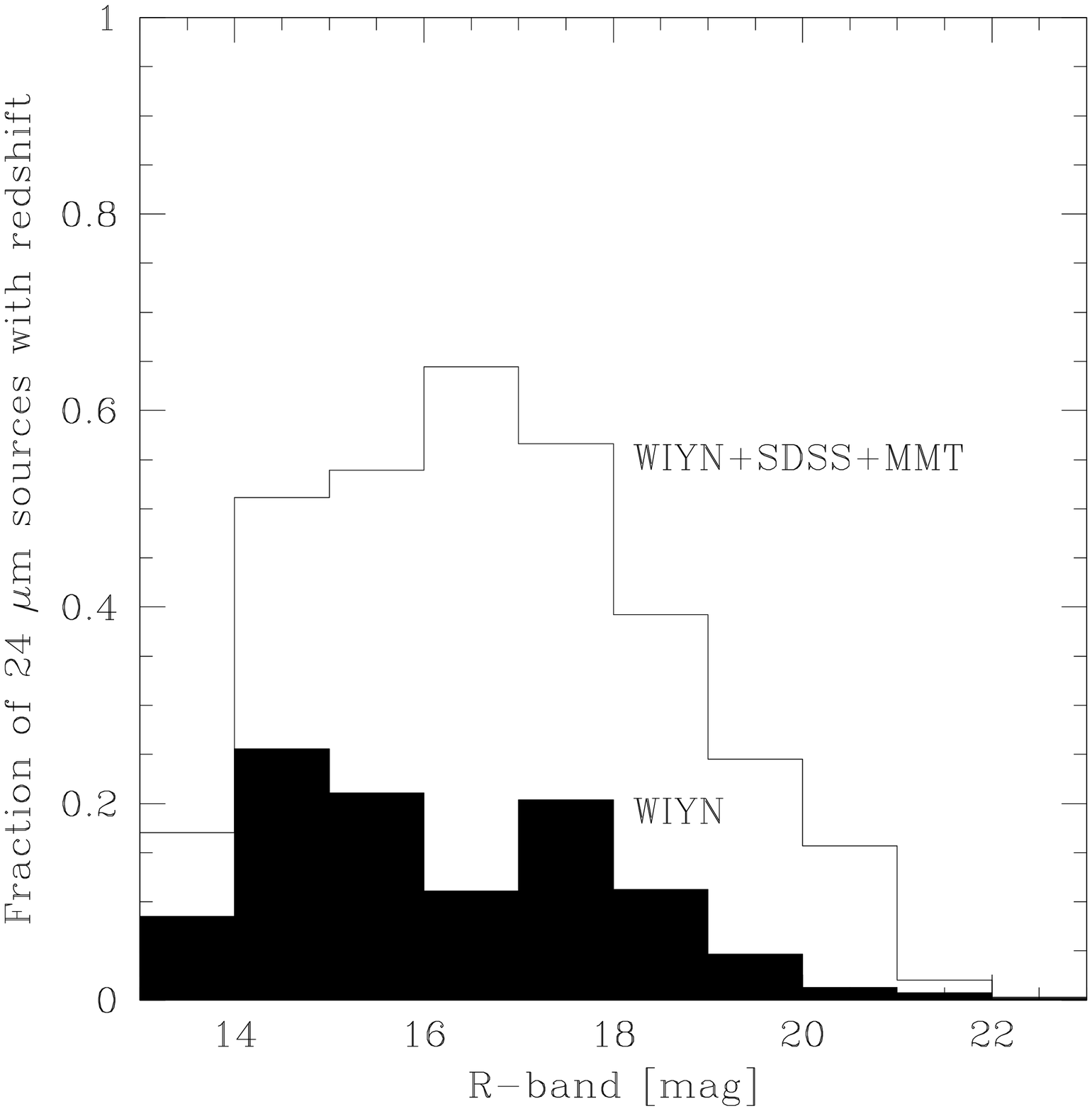}
}
\caption{\label{fig:zfrac24um} {\em Left}: Fraction of 24~\mum\
sources with spectroscopic redshifts in the FLS as a function of
24~\mum\ flux density. The stars were removed from the 24~\mum\
catalog. {\em Right}: Fraction of 24~\mum\ sources with redshift in
the FLS as a function of R-band magnitude. Our magnitude cutoff for
our survey was an R-band magnitude of 23.}
\end{figure}

\begin{figure}
\centerline{
\includegraphics[width=220pt,height=220pt,angle=0]{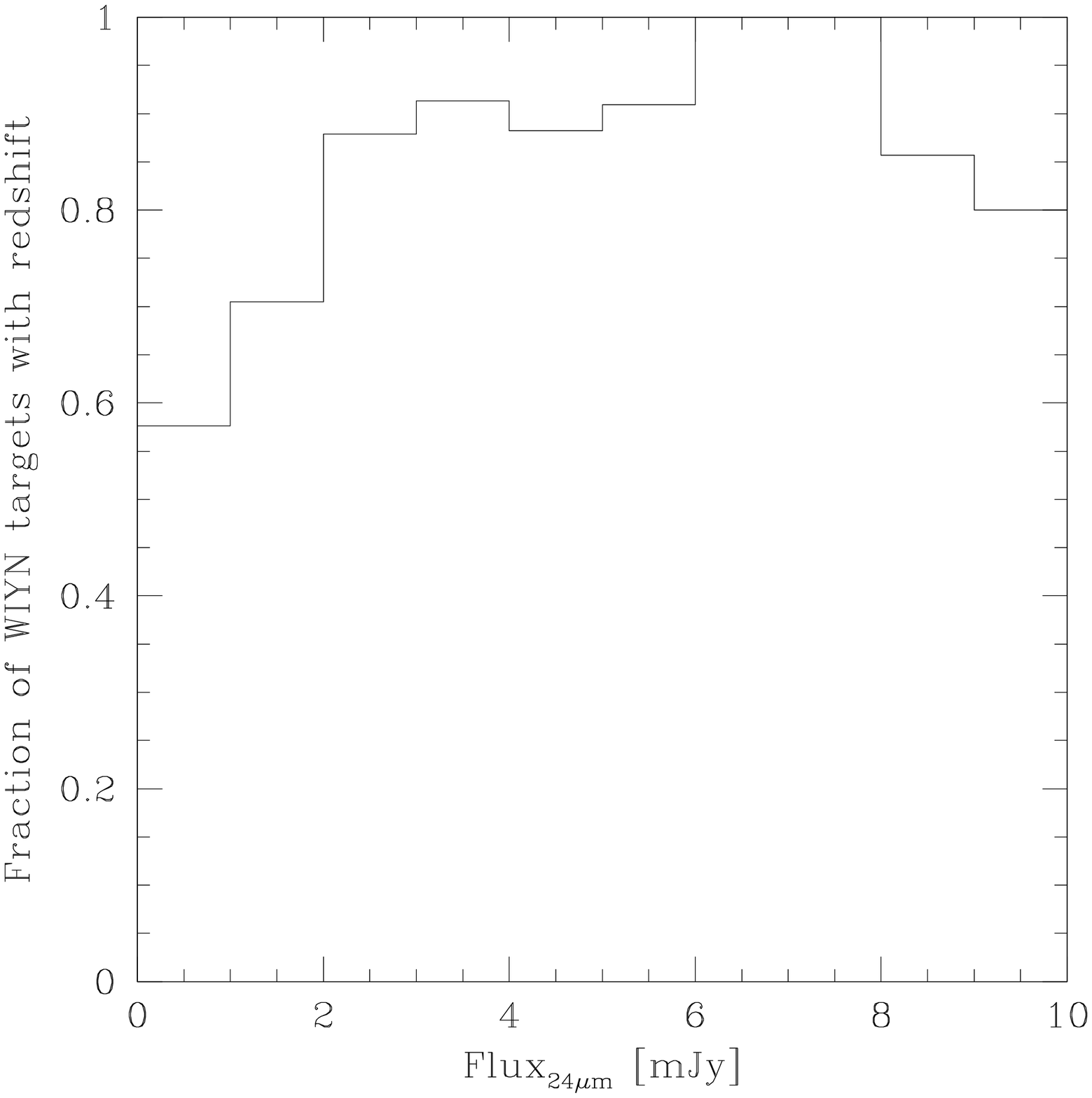}
\includegraphics[width=220pt,height=220pt,angle=0]{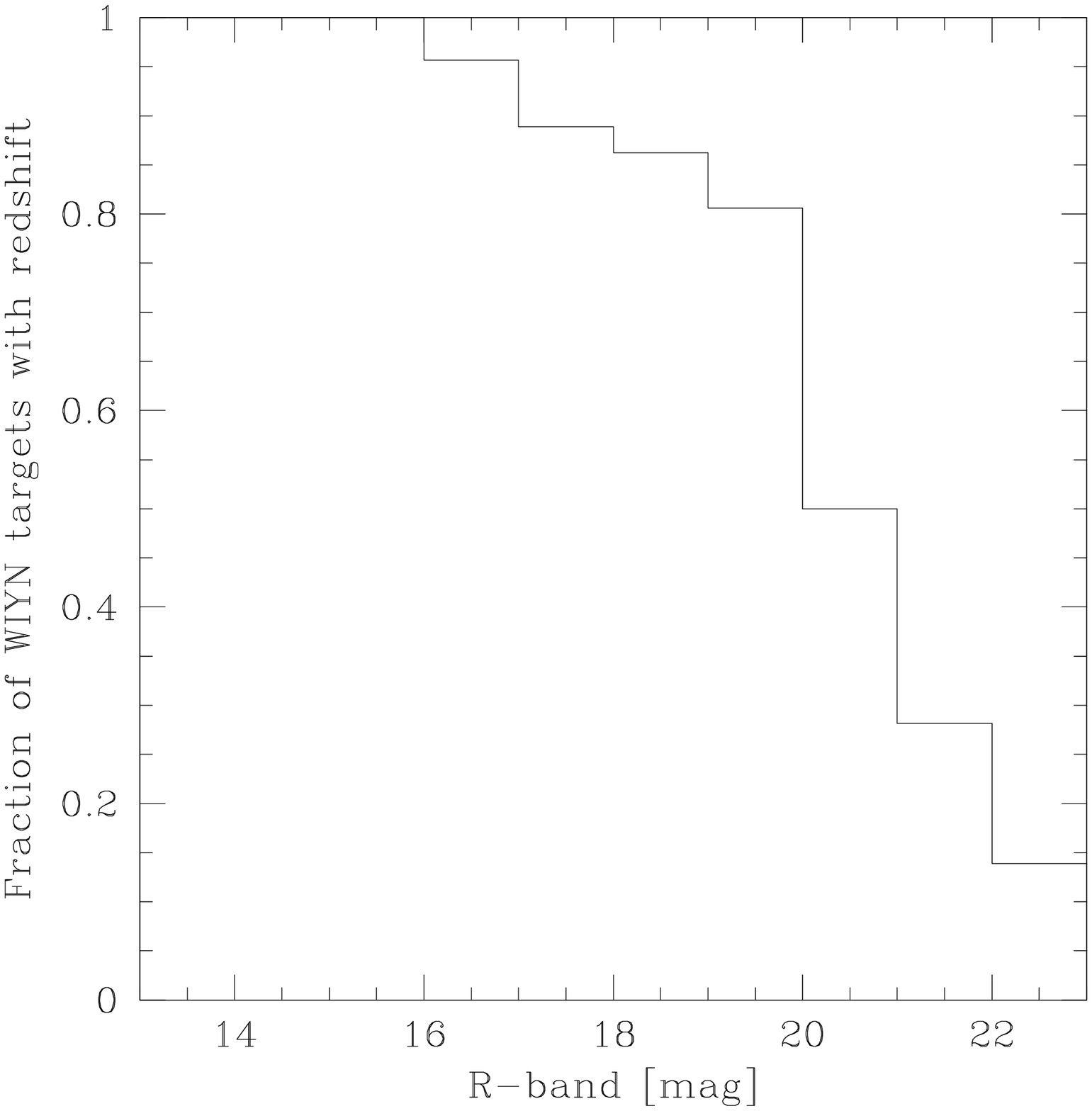}
}
\caption{\label{fig:zsuc} Fraction of WIYN/Hydra targets for which we
have successfully measured a redshift as a function of 24~\mum\ flux
density ({\em left}) and R-band magnitude ({\em right}).}
\end{figure}

\begin{figure}
\centerline{
\includegraphics[width=220pt,height=220pt,angle=0]{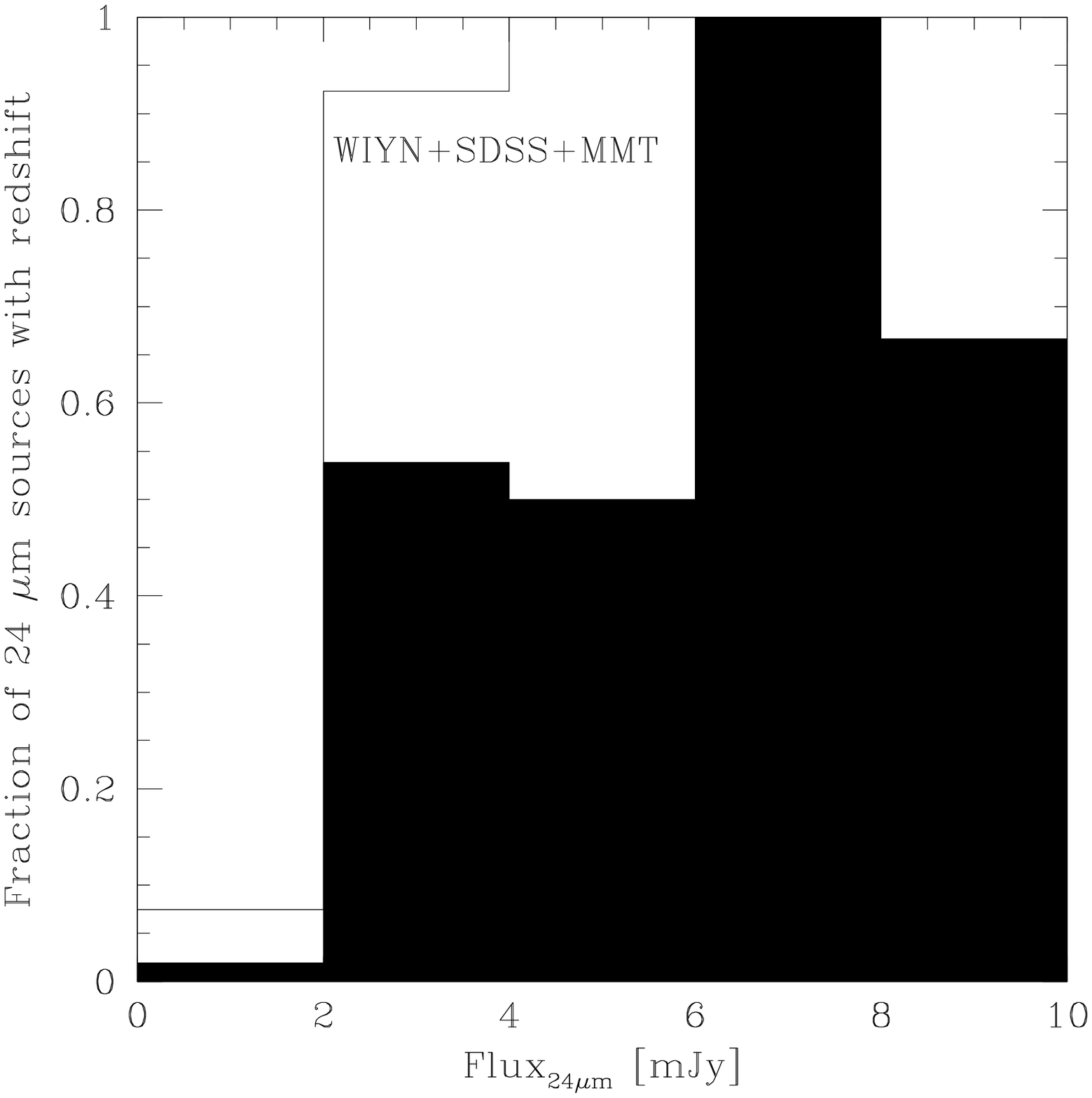}
\includegraphics[width=220pt,height=220pt,angle=0]{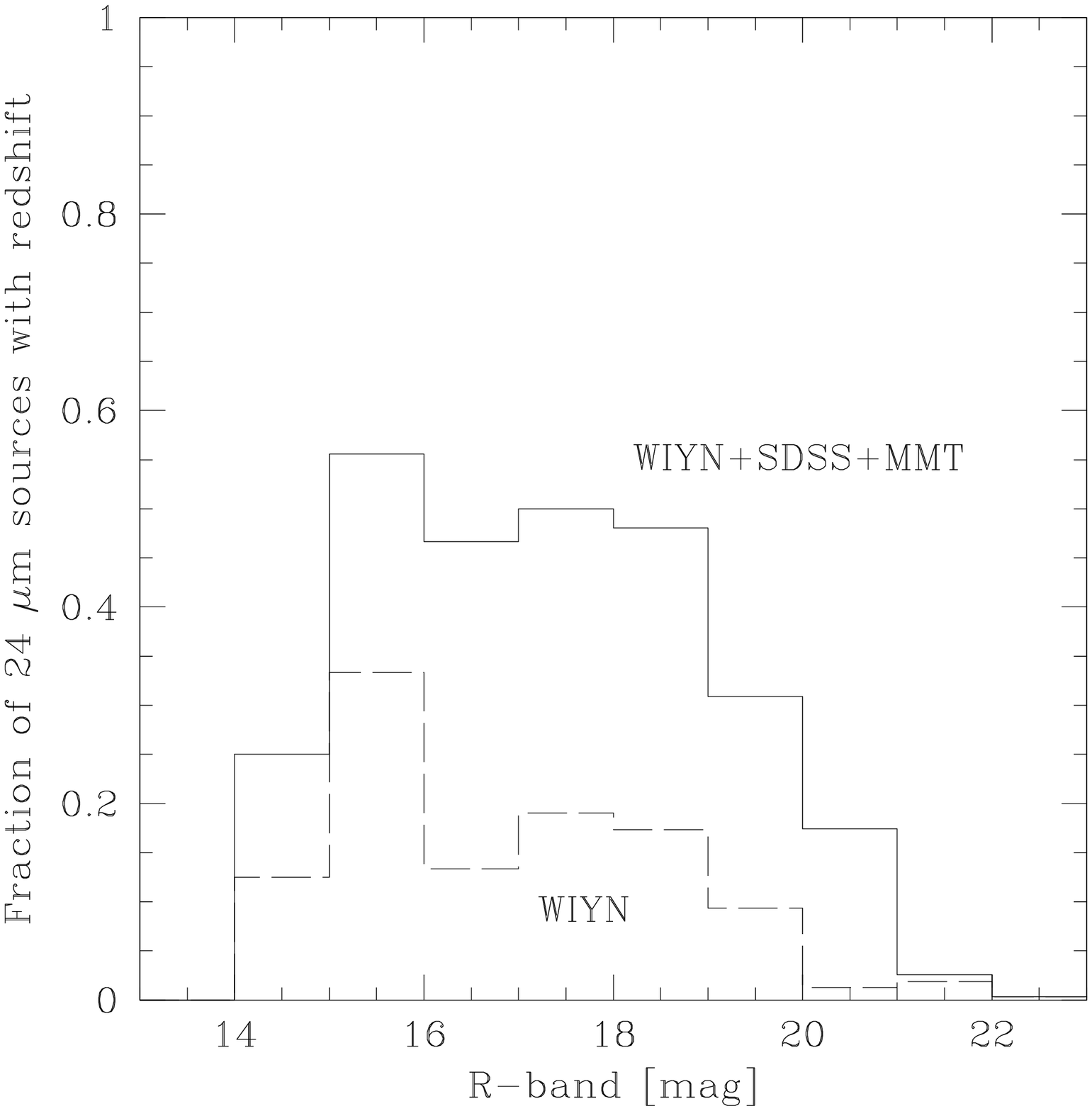}
}
\caption{\label{fig:zfrac24umver} Fraction of 24~\mum\ sources with
spectroscopic redshifts in the FLS verification region only, where our 
redshift survey is more complete, as a function of 24~\mum\ flux 
density ({\em left}) and R-band magnitude ({\em right}).}
\end{figure}

The fraction of 24~\mum\ sources for which we successfully obtained a
redshift is displayed in Figure~\ref{fig:zfrac24um}, as a function of
24~\mum\ flux density (left), and R-band magnitude (right). A redshift
was successfully obtained for 412 out of the 600 sources targeted with
WIYN/Hydra, i.e.\ with a success rate of $\sim$ 80\%. The success rate
is displayed in Figure~\ref{fig:zsuc}, as a function of 24~\mum\ flux
density (left), and R-band magnitude (right). As seen in
Figure~\ref{fig:fields}, more 24~\mum\ sources were targeted in the
deeper FLS verification (FLSV) region. The WIYN/Hydra redshift survey
is more complete there, with a success rate as shown in
Figure~\ref{fig:zfrac24umver}.

\section{Emission Line Measurements}

Each emission and absorption feature was identified and fitted with a
local continuum and a blend of Gaussian components, to determine line
centers, fluxes, equivalent widths (EWs), and the continuum
level. Errors were calculated using the raw counts. The uncorrected
line strengths and equivalent widths of the principal emission
features are given in an ASCII table, available in the electronic
edition of the {\it Astrophysical Journal}, with the format explained
in Table~\ref{tbl:lineem}.

\subsection{Extinction Correction}

The comparison of the observed and predicted Balmer line ratios can be
used to estimate the internal extinction of the galaxies in our
survey. The H$\alpha$ $\lambda$6563 and H$\beta$ $\lambda$4861 Balmer
lines, were detected in 134 WIYN/Hydra sources. Therefore, a direct
measurement of the Balmer decrement was possible for this subset. For
the other 317 emission-line sources, we assumed the median value of
our subset of 6.26. The color excess E(B-V)$_{gas}$ of each source was
computed by comparing the observed Balmer line ratio
(F$_o^{H\alpha}$/F$_o^{H\beta}$) with the intrinsic unobscured ratio
using the equation

\begin{equation}
E(B-V)_{gas} = \frac{2.5}{[k(H\alpha)-k(H\beta)]} 
\log \left[ \frac{F_o^{H\alpha}/F_o^{H\beta}}{F_i^{H\alpha}/F_i^{H\beta}} \right].
\end{equation}

\noindent Here, the intrinsic unobscured line ratio
F$_i^{H\alpha}$/F$_i^{H\beta}$ was set equal to 2.87, assuming case B
recombination and T=10,000 K (Osterbrock 1989). The reddening curve
$[k(\lambda)]$ was taken from Calzetti (2001), for starburst
galaxies, and is valid between 0.12 to 2.2~\mum. The color excess can
also be converted into a wavelength-dependent extinction based on the
relation

\begin{equation}
A(\lambda) = E(B-V)_{gas} \; k(\lambda),
\end{equation}

\noindent where $A(\lambda)$ is the mean emission-line-derived
extinction in units of magnitudes at the wavelength $\lambda$.

Using the computed color excess, all emission-lines were dereddened
for each galaxy using the routine {\em calz\_unred}~\footnote{A value
of R$_V$=4.05 was assumed.} from the IDL astronomical library. Given
the uncorrected line fluxes and color excesses, this code returns the
dereddened line fluxes. The Balmer decrement, color excess and
extinction, A(V), calculated for each source in our sample, can be
found in an ASCII table, available in the electronic edition of the
{\it Astrophysical Journal}, with the format explained in
Table~\ref{tbl:lineext}.

\subsection{Diagnostic Line Ratios} 

Diagnostic line-intensity ratios, such as [OIII]/H$\beta$ and
[NII]/H$\alpha$, corrected for reddening, are effective at separating
populations with different ionization sources (Osterbrock 1989; Kewley
et al.\ 2001). The ionizing radiation field found in active galaxies
is harder than in star-forming galaxies, and this gives higher values
of [NII]/H$\alpha$ and [OIII]/H$\beta$. Hence, the narrow line AGNs
are found in the upper right portion of the [OIII]/H$\beta$ versus
[NII]/H$\alpha$ diagnostic diagram. The lines in these ratios are also
selected to be close in wavelength space in order to minimize the
effect of dust extinction on the computed line ratios.

\begin{figure}
\centerline{
\includegraphics[width=240pt,height=240pt,angle=0]{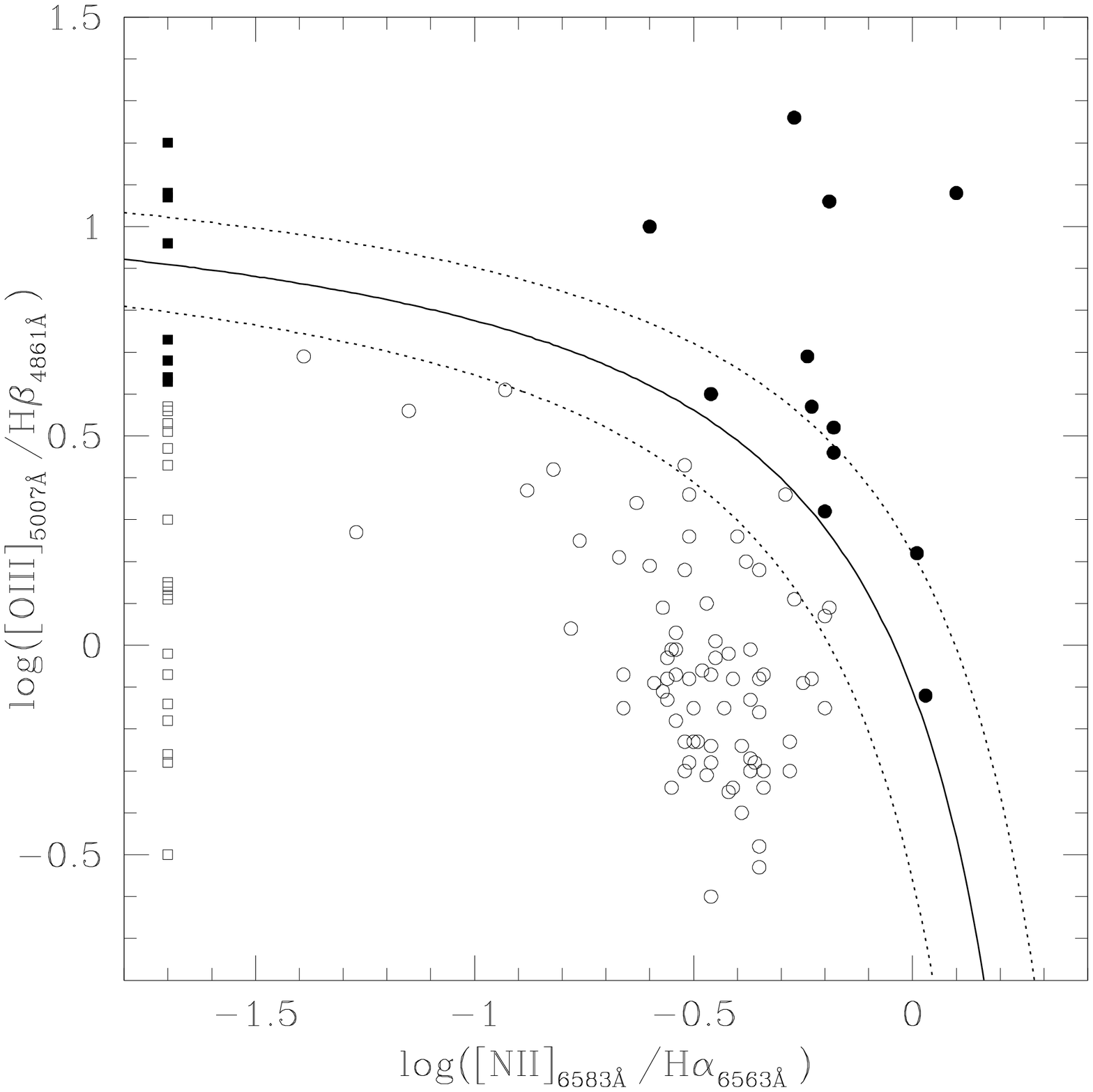}
\includegraphics[width=240pt,height=240pt,angle=0]{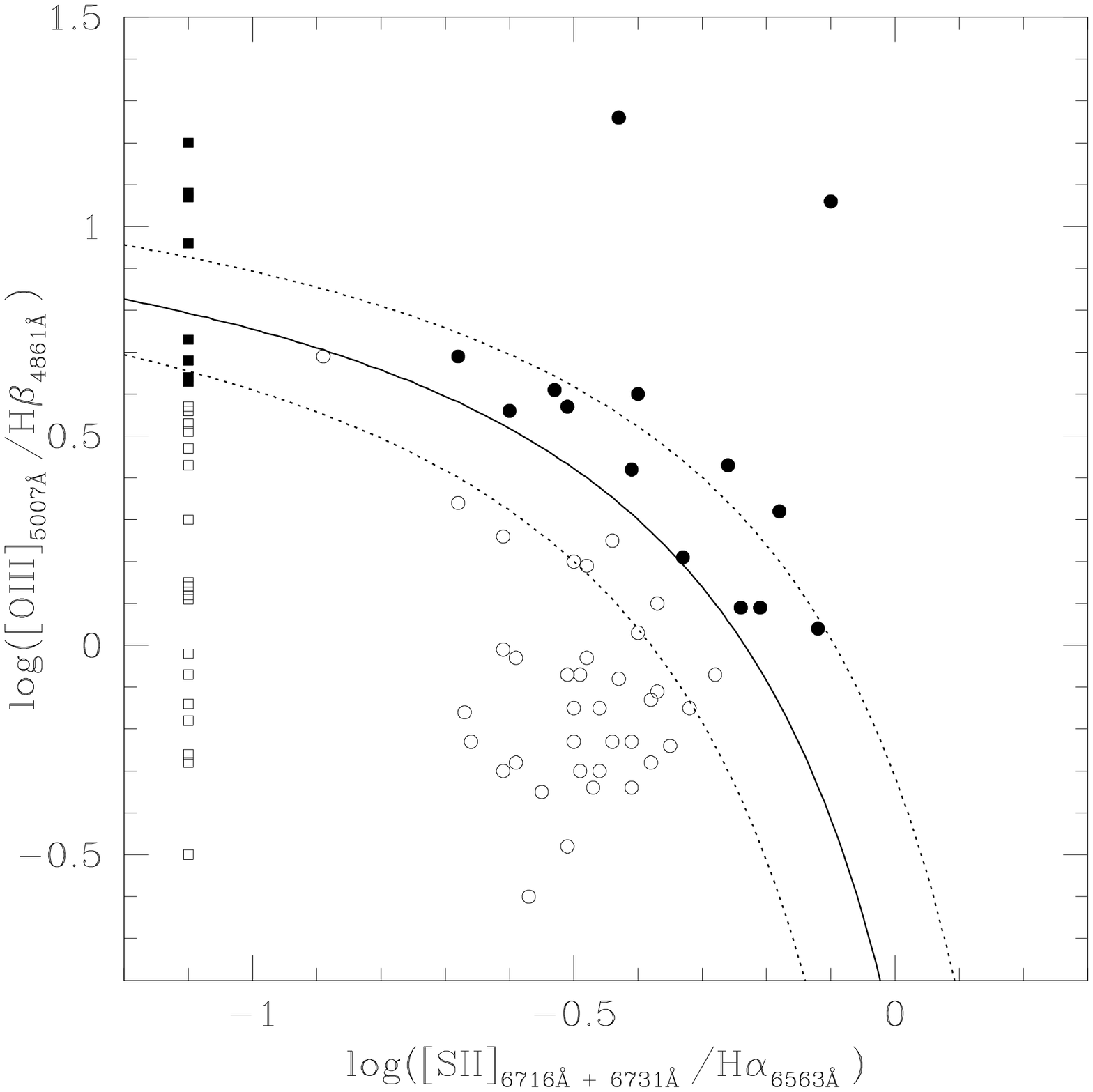}
}
\caption{\label{fig:diag} [OIII]/H$\beta$ vs. [NII]/H$\alpha$
diagnostic diagram ({\em left}) and [OIII]/H$\beta$
vs. [SII]/H$\alpha$ ({\em right}), for the WIYN/Hydra spectroscopic
sources where these emission lines were detected. HII region-like objects are
plotted with {\em open circles}, whereas objects classified as narrow
line AGNs are shown as {\em filled circles}. The {\em solid line} is
the theoretical starburst-AGN classification line from Kewley et
al. (2001). The {\em dotted line} shows the error range of the Kewley
et al.\ model predictions ($\pm 0.1$ dex range of {\em solid line}). }
\end{figure}

We were able to identify narrow line AGNs in our WIYN/Hydra
spectroscopic sample using diagnostic diagrams of [OIII]/H$\beta$
versus [NII]/H$\alpha$ and [OIII]/H$\beta$ versus [SII]/H$\alpha$. The
values of these line ratios are given in an ASCII table, available in
the electronic edition of the {\it Astrophysical Journal}, with the
format explained in Table~\ref{tbl:lineext}. The results are displayed
in Figure~\ref{fig:diag}. At first glance, this figure reveals that
the majority of emission-line sources in this subsample are
star-forming systems. The theoretical starburst-AGN classification,
shown in Figure~\ref{fig:diag} as the solid curve, is a result of the
modeling of starburst galaxies of Kewley et al.\ (2001). These models
are generated using the ionizing ultraviolet radiation fields produced
by stellar population synthesis models (in this case PEGASE v2.0), in
conjunction with a detailed self-consistent photoionization model
(MAPPINGS III). Using the Kewley et al.\ classification, we identified
twenty sources in this subsample as narrow line AGNs (shown as filled
circles in Figure~\ref{fig:diag}).

Sources without a [NII]/H$\beta$ flux ratio measurement were
classified as AGNs when their [OIII]/H$\beta$ flux ratio was greater
than 0.6 (shown as filled squares in Figure~\ref{fig:diag}). This
value was chosen based on the Kewley et al.\ curve and the median
value of [NII]/H$\alpha$ for the subsample, i.e.\ -0.5. This is the
most conservative limit of the two diagnostic diagrams. Among these
sources, another eight narrow line AGNs were found, bringing the total
to 28. Adding the eight broad emission line AGNs, we were able to
identify, based on the properties of the optical spectra, a total of
36 AGNs in the WIYN/Hydra catalog. Identification of AGNs in the SDSS
and MMT/Hectospec catalog was done using the {\em spec\_class} and
{\em class} index, respectively. In total, 125 and 298 AGNs were
flagged in the radio and infrared catalog, respectively.

\subsection{H$\alpha$ and Radio Star Formation Rates}

\begin{figure}
\centerline{
\includegraphics[width=240pt,height=240pt,angle=0]{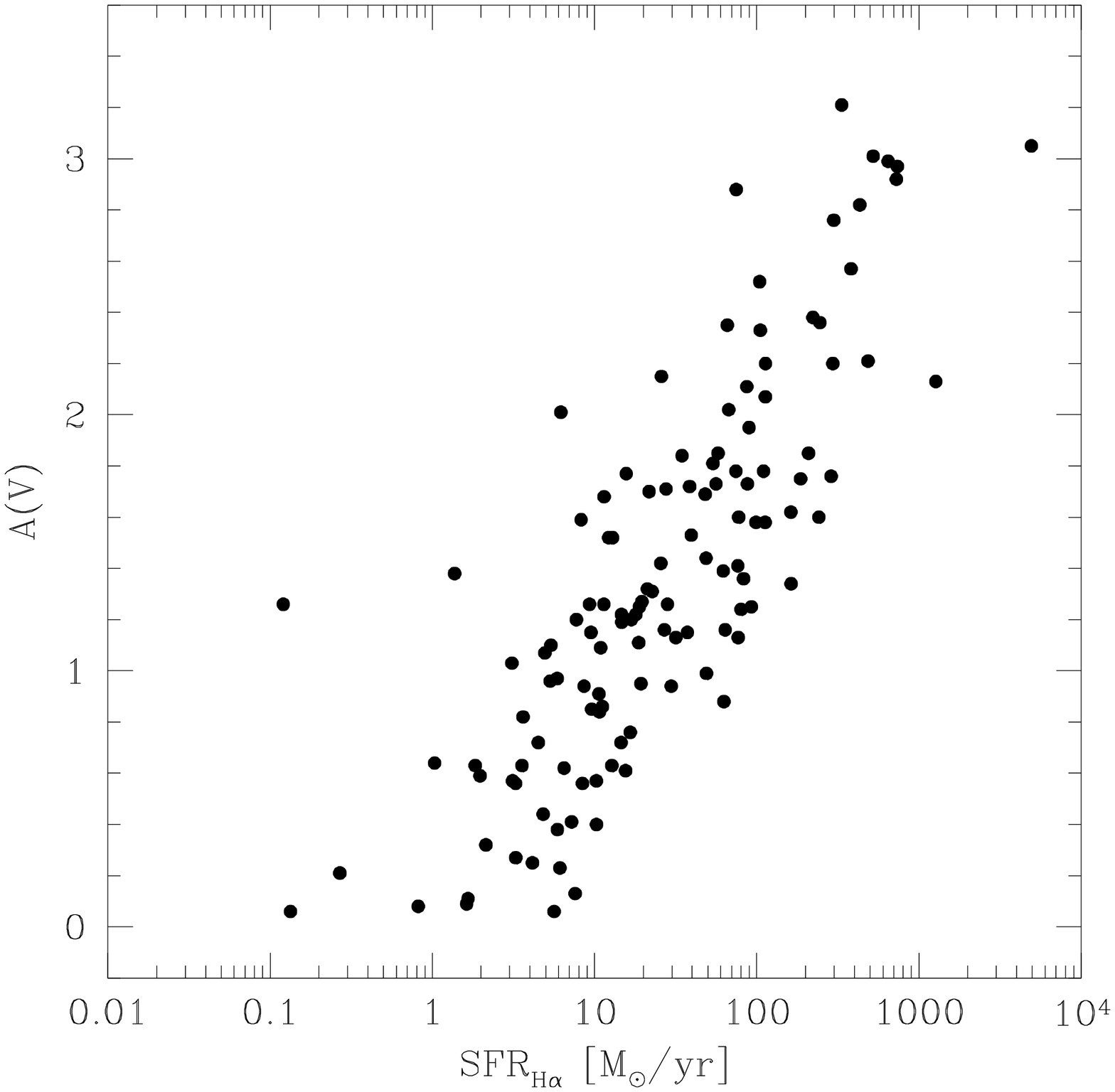}
\includegraphics[width=240pt,height=240pt,angle=0]{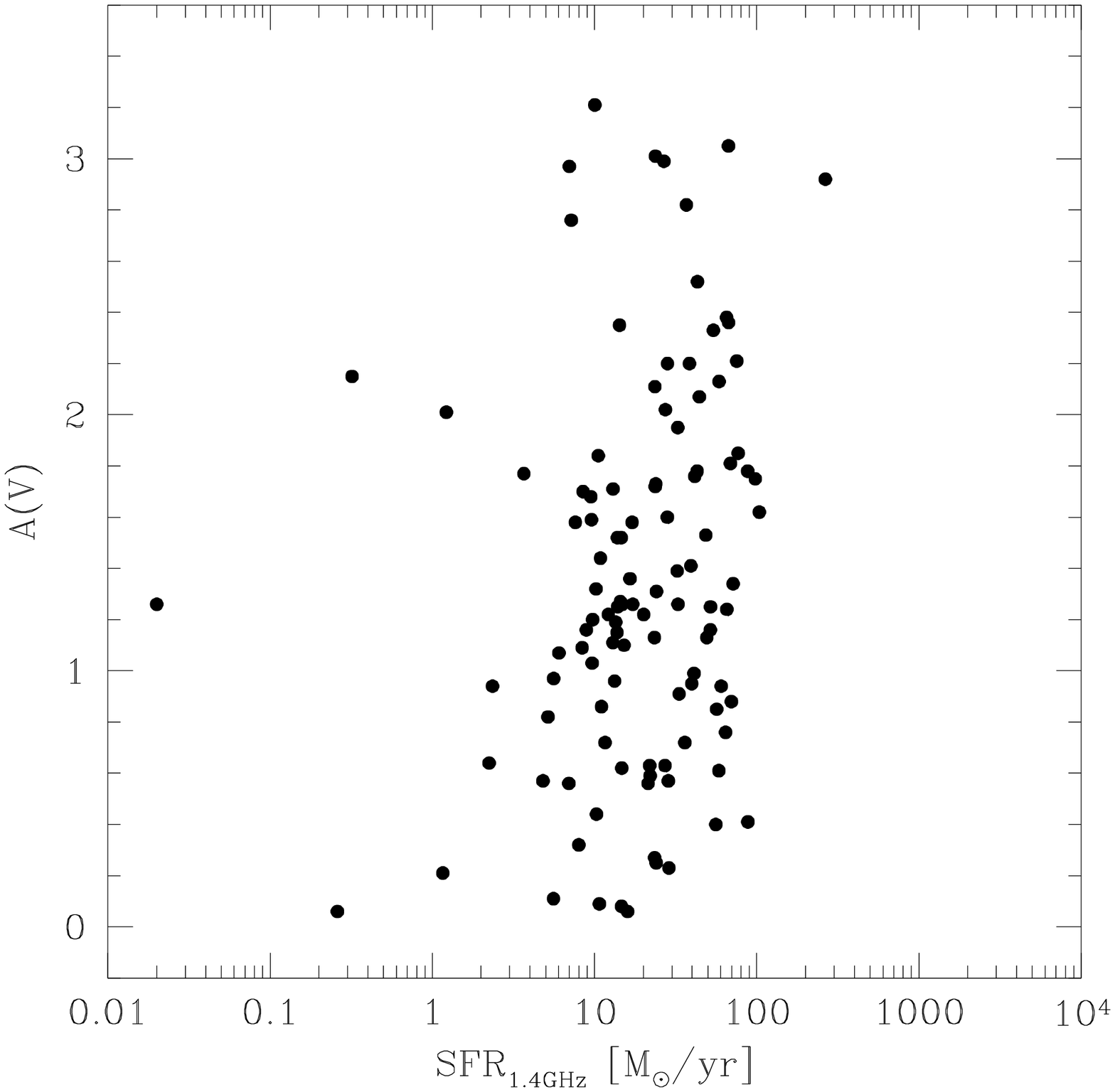}
}
\caption{\label{fig:SFR} {\em Left}: Optical extinction versus SFR
measured from H$\alpha$, for the WIYN/Hydra sources with measured
H$\alpha$ and H$\beta$ fluxes. The AGNs, as selected from the
diagnostic line ratios, were removed. {\em Right}: Optical extinction
versus SFR estimated from the radio luminosity.}
\end{figure}

We calculated the star formation rates (SFRs) for the 254 sources in
our survey with detected H$\alpha$ line emission (corrected for
extinction). We used the SDSS r' images to estimate the
aperture correction needed to transform our measured H$\alpha$ fluxes,
obtained using 2 arcsec diameter fibers, to an integrated emission-line
flux. The measured H$\alpha$ fluxes from HII regions can be converted
into total SFRs using the H$\alpha$-SFR calibration relation derived
by Kennicutt (1998)

\begin{equation}
SFR_{H\alpha} = 7.9\times10^{-42} L_{H\alpha},
\end{equation}

\noindent where $L_{H\alpha}$ is the measured line luminosity in ergs
s$^{-1}$ and the SFR$_{H\alpha}$ is given in M$_\odot$ yr$^{-1}$. The
conversion between ionizing flux and the star formation rate generally
applies only to OB stars with masses of $> 10$ M$_\odot$ and lifetimes
$< 20$ Myr. However, the conversion factor in equation~(3) was
extrapolated to a total SFR assuming solar abundances and a Salpeter
initial mass function (IMF) over the mass range 0.1 $<$ M/M$_\odot <$
100. The main limitations of the method are its sensivity to
uncertainties in extinction and the IMF (a change from Salpeter to
Scalo IMF will increase the SFR by a factor of 3), and the assumption
that all of the massive star formation, which contribute most to the
integrated ionized flux, is traced by ionized gas.  The calculated
values of SFR$_{H\alpha}$ are given in an ASCII table, available in
the electronic edition of the {\it Astrophysical Journal}, with the
format explained in Table~\ref{tbl:lineext}. Note that the sources
classified as AGNs were removed from this sample, as the Kennicutt
relation does not apply to these systems. We estimated star formation
rates in the range $0.2 - 200$ M$_\odot$ yr$^{-1}$, and found that
these values were strongly correlated with extinction (see
Figure~\ref{fig:SFR}, left).

The star formation rates can also be estimated from the radio
luminosity of galaxies (see Condon 1992, equation~21 and
23). Following equation~(28) of Condon, Cotton \& Broderick (2002), we
can write

\begin{equation}
SFR_{1.4~GHz} = 1.2\times10^{-21} L_{1.4~GHz},
\end{equation}

\noindent where $L_{1.4~GHz}$ is the measured radio luminosity in W
Hz$^{-1}$, and the SFR$_{1.4~GHz}$ is given in M$_\odot$
yr$^{-1}$. The more massive stars in normal galaxies are responsible
for most of the radio emission. Given that their lifetime is shorter
than a Hubble time, the current radio luminosity is proportional to
the recent star formation rate. The global radio non-thermal
(synchrotron radiation) and thermal (free-free emission) luminosities,
as well as the FIR/radio ratio, can be computed assuming an average
star formation rate of stars more massive than 5 M$_\odot$. The star
formation rate is then extrapolated to stars with M $>$ 0.1 M$_\odot$,
by assuming a Salpeter IMF over the mass range 0.1 $<$ M/M$_\odot <$
100. The calculated values of SFR$_{1.4~GHz}$ can also be found in an
ASCII table, available in the electronic edition of the {\it
Astrophysical Journal}, with the format explained in
Table~\ref{tbl:lineext}. As can be seen in Figure~\ref{fig:SFR}
(right), the SFR$_{1.4~GHz}$ does not correlate with optical
extinction.

\begin{figure}
\centerline{
\includegraphics[width=240pt,height=240pt,angle=0]{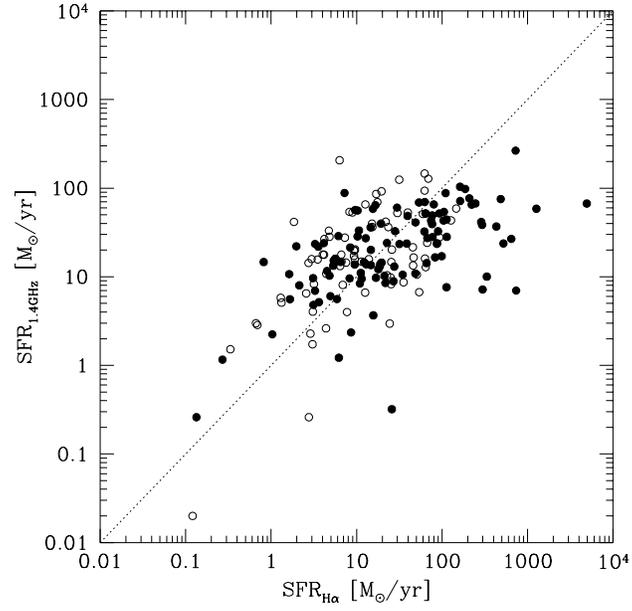}
}
\caption{\label{fig:SFRrat} Comparison between the star formation
rates computed from the H$\alpha$ and 1.4~GHz luminosities. All sources
with dereddened H$\alpha$ fluxes and detected at 1.4~GHz are shown as
{\em open circles}. The {\em filled circles} designate the sources for
which a Balmer decrement was measured directly (otherwise, the median
value of 6.26 was assumed). The {\em dotted line} is the linear
relationship with zero intercept and slope equals to one.}
\end{figure}

\begin{figure}
\centerline{
\includegraphics[width=240pt,height=240pt,angle=0]{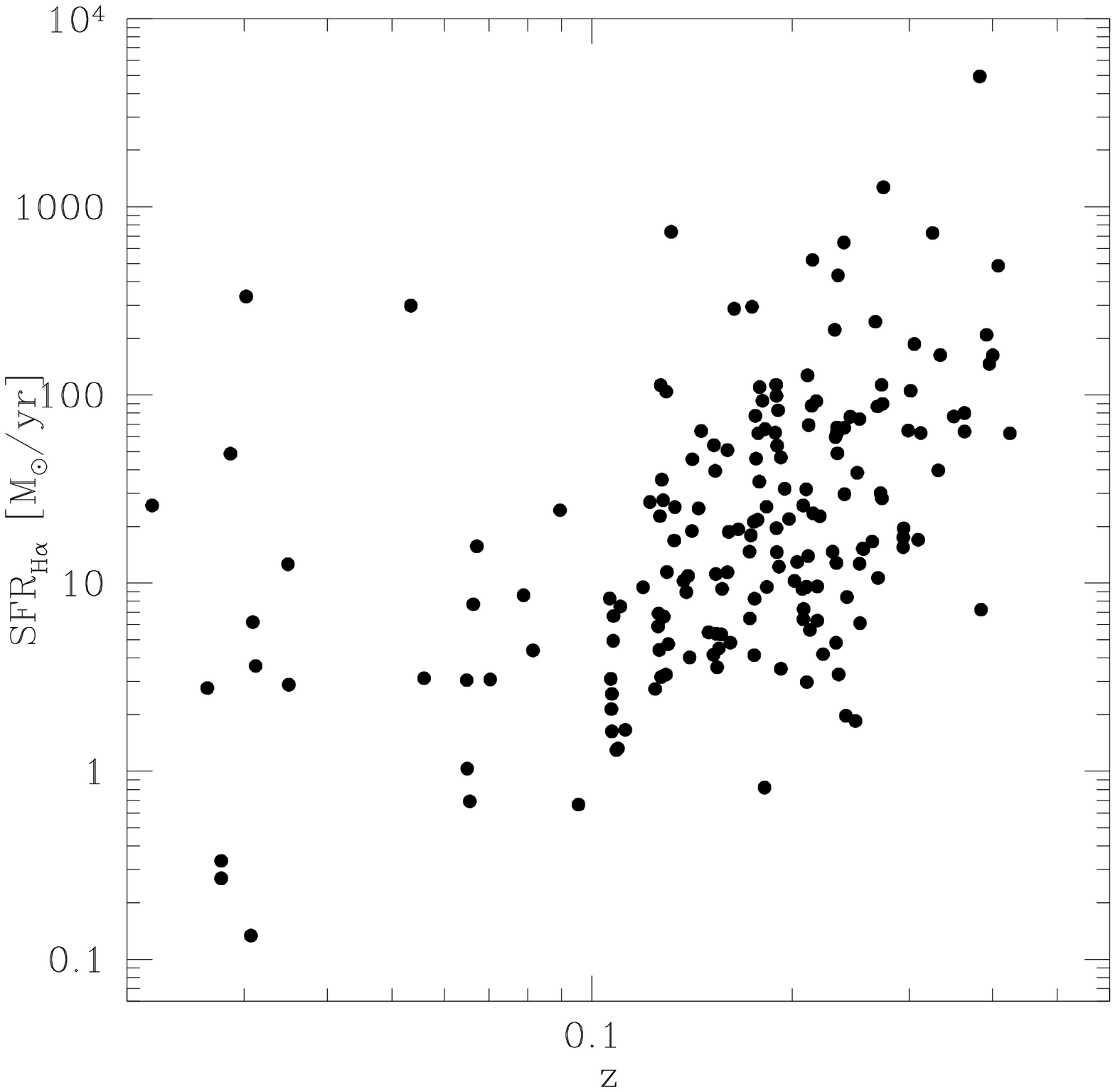}
\includegraphics[width=240pt,height=240pt,angle=0]{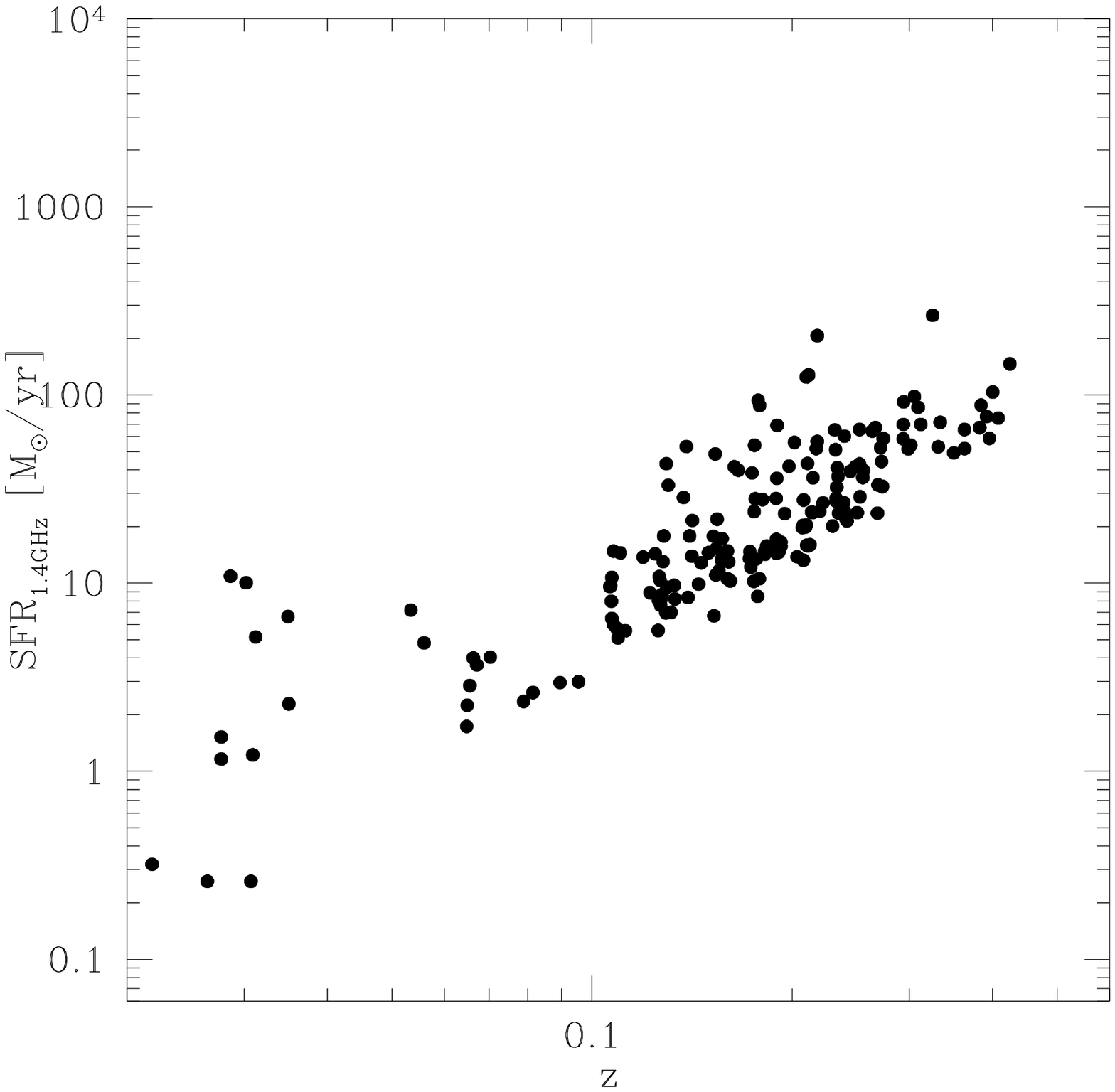}
}
\caption{\label{fig:SFRz} Star formation rates computed from 
H$\alpha$ ({\em left}) and 1.4~GHz luminosities ({\em right}) 
as a function of redshift.}
\end{figure}

The star formation rates derived from the H$\alpha$ and 1.4~GHz
luminosities are compared in Figure~\ref{fig:SFRrat}. The star
formation rate computed from the radio luminosity is on average
consistent with the one derived from the H$\alpha$ optical line
emission. The star formation rates start to deviate slightly from this
simple correlation at the upper end of the SFR range, as the optical
extinction increases (see Figure~\ref{fig:SFR}, left), and therefore
the H$\beta$ flux measurement becomes more uncertain. Moreover, we
don't necessarily expect a perfect agreement between the two methods
as the current radio emission is proportional to the rate of massive
star formation occuring during the past 100~Myr, whereas the H$\alpha$
line emission is associated with more recent, i.e.\ $< 20$~Myr,
star-forming activity. The star formation rates are plotted as a
function of redshift in Figure~\ref{fig:SFRz}. At low redshift ($z <
0.4$), our survey is detecting sources with a wide range of
star-forming activity. However, at higher redshifts, we are only
detecting sources with a SFR $> 40$ M$_\odot$ yr$^{-1}$.

\section{Luminosities and Infrared-Radio Correlation}

Radio and infrared luminosities were calculated for a total of 830
radio sources and 1753 24~\mum\ sources. The observed radio flux
densities were corrected to the rest-frame flux densities by applying
a multiplicative factor of $(1+z)^{0.7}$, assuming a power law
emission of $S \propto \nu^{-0.7}$ at these frequencies. The flux
densities were also corrected for the effect of bandwidth compression
by applying the multiplicative term of $(1+z)^{-1}$.

The k-correction for the 24~\mum\ passband was computed using
redshifted spectral energy distribution (SED) templates. The majority
of 24~\mum\ sources are star forming galaxies, but their SEDs are not
known a priori. However, if fluxes are also measured at other
wavelengths, such as 4.5, 24 and 70~\mum, it becomes possible to
select a best fit SED from a database of SEDs, such as the one of
Siebenmorgen \& Kr\"ugel (2007), and compute a correction specific to
each source. To see if we could apply this technique, we used the
70~\mum\ source catalogs of Frayer et al.\ (2004). Unfortunately, only
22\% of all sources with measured redshifts were detected at
70~\mum. With most of the 24~\mum\ population unknown, and with
luminosities in the range of normal star-forming galaxies, we
therefore elected to use the SED of the typical starburst galaxy M82
as our template for calculating the k-correction (data from Sturm et
al.\ 2000). We obtained identical results when using the average IRS
spectrum of 13 starburst galaxies (Brandl et al.\ 2006). As can be
seen in Figure~6 of Brandl et al.\ (2006), this is not surprising, as
this average IRS spectrum and the M82 SED are virtually the same at
these wavelengths.

However, extreme sources, such as high redshift luminous infrared
galaxies (LIRGs), ultra-luminous infrared galaxies (ULIRGs) and AGNs,
are not expected to be well represented by this template. To estimate
the error on the luminosity calculations at the bright end of the
luminosity function of star forming galaxies, we re-computed the
k-correction using two additional templates: UGC~5101 (data from Armus
et al.\ 2007), a typical ULIRG, and Arp~220 (data from Sturm et al.\
1996), one of the most extreme objects of this type (see Figure~2 of
Armus et al.\ 2007). The 24~\mum\ luminosities computed using the M82
and Arp~220 templates differed by $< 2$~\%, over the redshift range of
our survey. The difference was even less when comparing to our typical
ULIRG, UGC 5101. Similarly, the impact on the measurement of q$_{24}$
(see below) was found to be insignificant; using Arp~220, we found a
difference of 0.8~\% in the median and 0.2~\% in the dispersion. The
use of M82 for the k-correction was therefore found to be completely
adequate for our sample. Moreover, this result is consistent with the
work of Appleton et al.\ (2004), who performed a comparison between
the k-correction computed from best fit model SEDs and the SED of
M82. We stress, however, that this assumption is likely to break down
at higher redshifts, i.e.\ $z > 1.2$, when differences in the SEDs
start to have a large impact on the 24~\mum\ flux densities. As for
the AGNs identified in our sample, the Circinus galaxy, a Seyfert 2
object, was assumed as a typical template and used to compute the
k-correction (data from Sturm et al.\ 2000).

\begin{figure}
\centerline{
\includegraphics[width=240pt,height=240pt,angle=0]{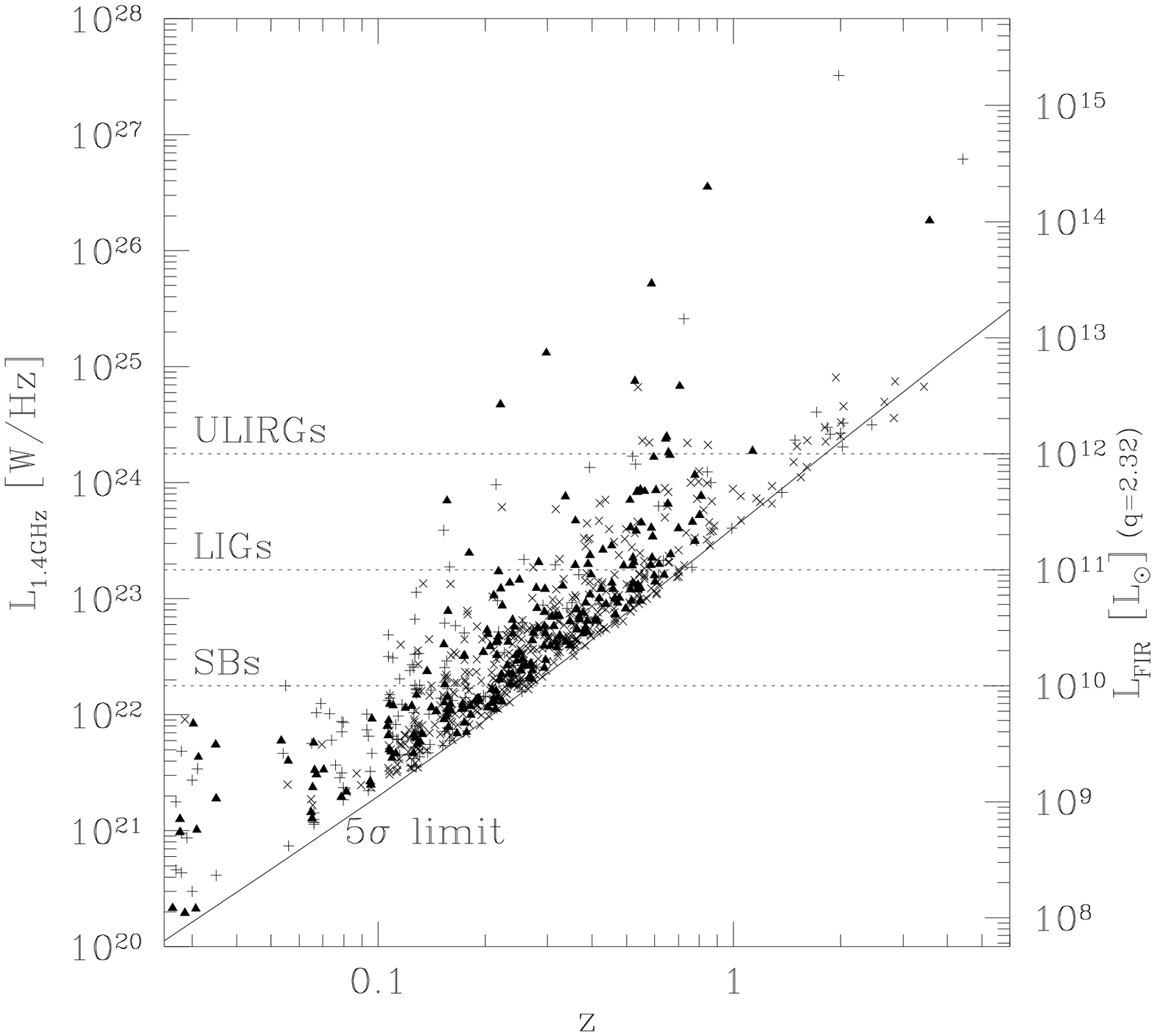}
\includegraphics[width=240pt,height=240pt,angle=0]{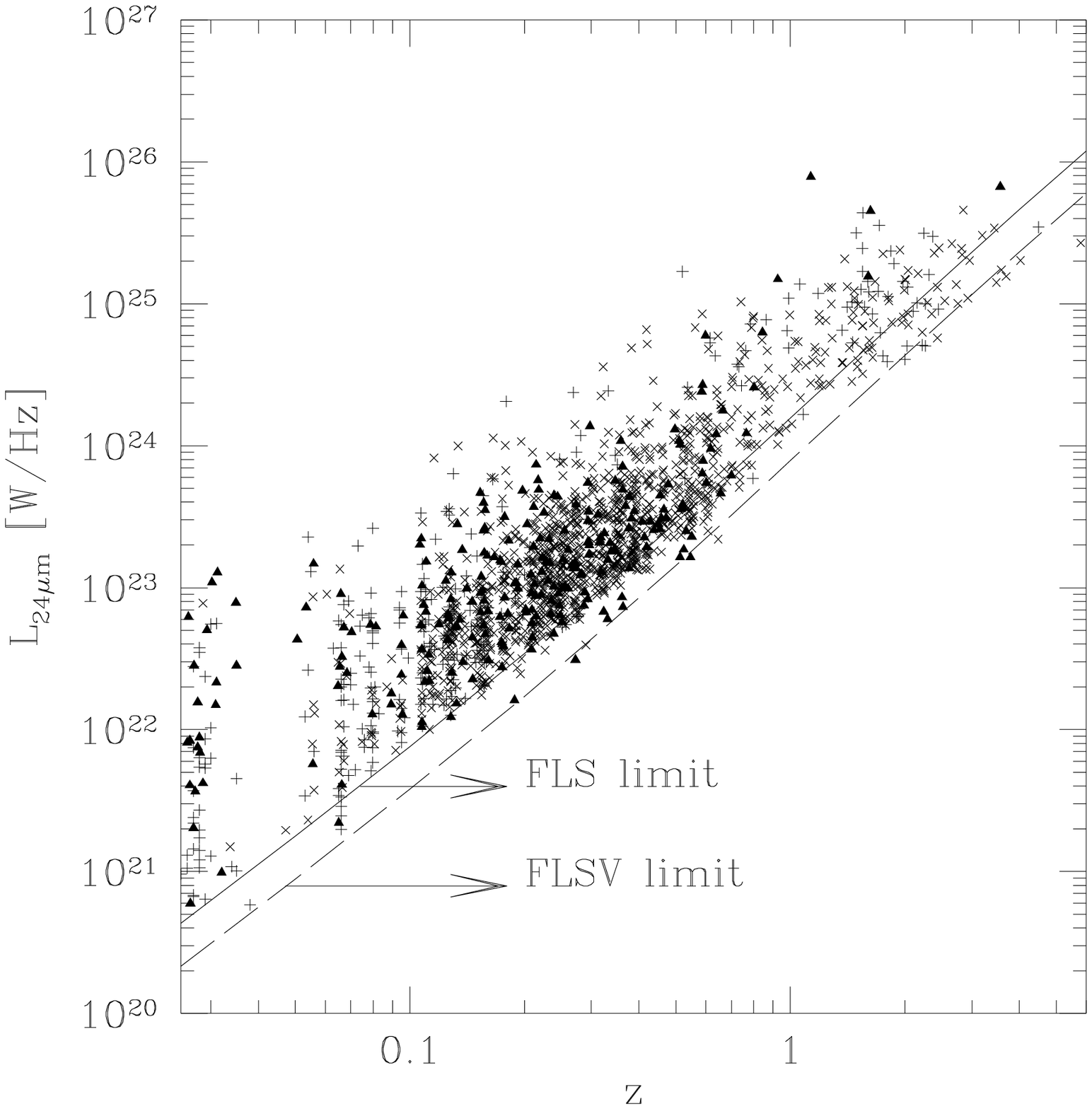}
}
\caption{\label{fig:lum} {\em Left}: The distribution of radio
luminosities with redshift in the {\em Spitzer} FLS region for SDSS
({\em crosses}), WIYN ({\em filled triangles}) and MMT ({\em diagonal
crosses}), from our 1.4~GHz and 24~\mum-selected sample. For
comparison, we show the relative scaling between the radio and FIR
flux densities (left- and right-hand scale, respectively), assuming a
value of q=2.32 for the slope of the radio-FIR correlation (Helou et
al.\ 1985). This value is only appropriate for star-formation
dominated systems. The solid line shows the 5$\sigma$ flux density
limit of the radio survey. {\em Right}: The distribution of 24~\mum\
luminosities with redshift. The 50\% completeness limit of the FLS
({\em solid line}) and FLSV ({\em long-dash line}) are also shown.}
\end{figure}

The luminosity distribution of the radio sources from our WIYN survey
is shown as a function of redshift in Figure~\ref{fig:lum} (left). The
majority of galaxies detected in the redshift range $z = 0 - 0.4$ have
low to medium radio luminosities, typical of star-forming galaxies,
whereas almost all the higher redshift galaxies are of higher radio
luminosity, comparable with powerful luminous infrared galaxies and
radio-quiet quasars. The 0.09~mJy flux density limit of the radio
survey (solid line in Figure~\ref{fig:lum}, left) tends to truncate
the shape of the luminosity function at $z > 0.3$. The 24~\mum\
luminosity distribution is also shown in Figure~\ref{fig:lum} (right),
along with its flux density limits (50\% completeness limits) of
0.3~mJy, for the FLS main (FLS) data, and 0.15~mJy, for the deeper FLS
verification (FLSV) observations. We plotted the optical extinction as
a function of 1.4~GHz and 24~\mum\ luminosity for the 134 WIYN/Hydra
sources with measured H$\alpha$ and H$\beta$ fluxes. No obvious trend
was seen. The optical extinction does not appear to correlate with
redshift either.

\begin{figure}
\centerline{
\includegraphics[width=300pt,height=300pt,angle=0]{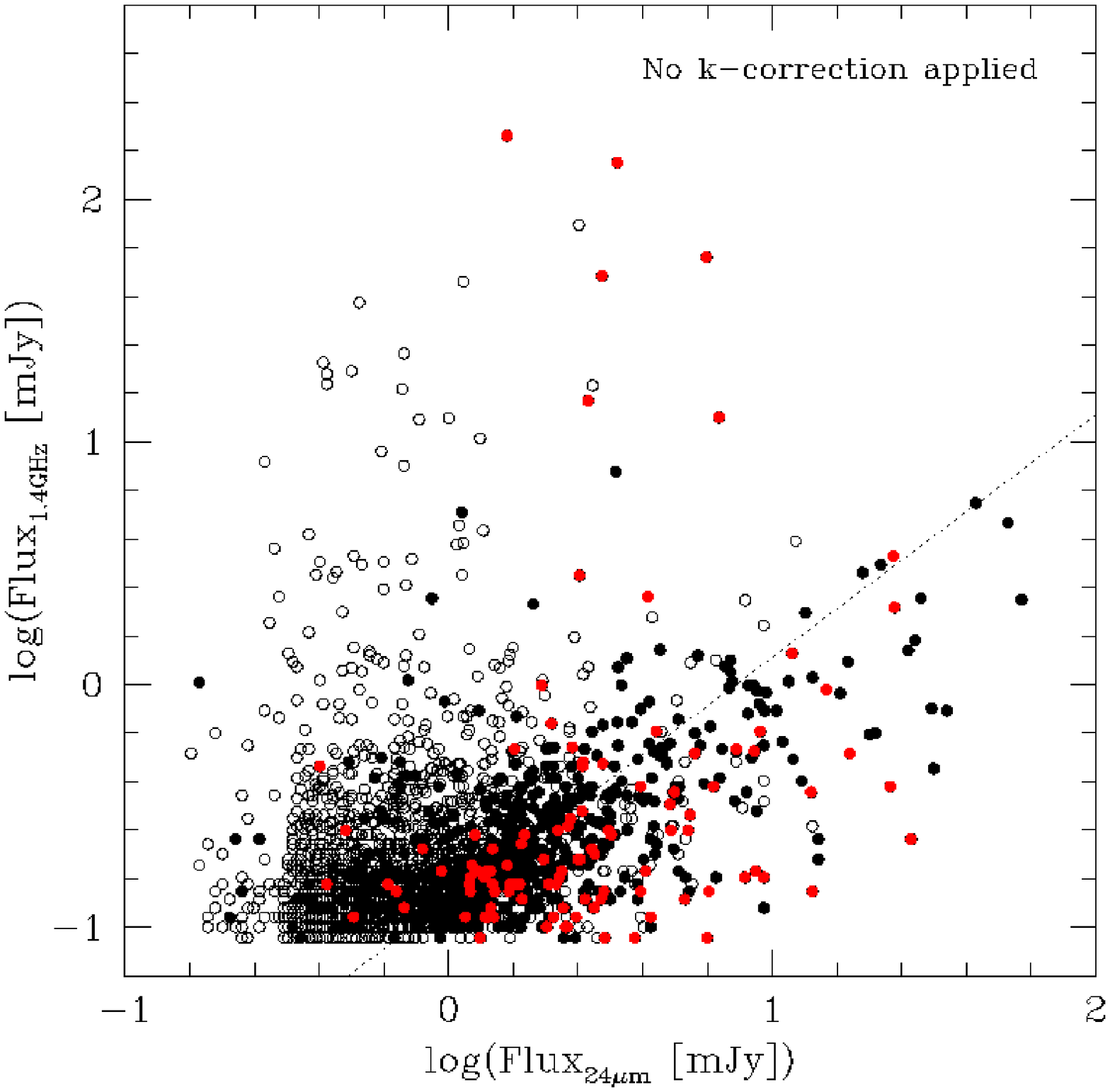}
\includegraphics[width=300pt,height=300pt,angle=0]{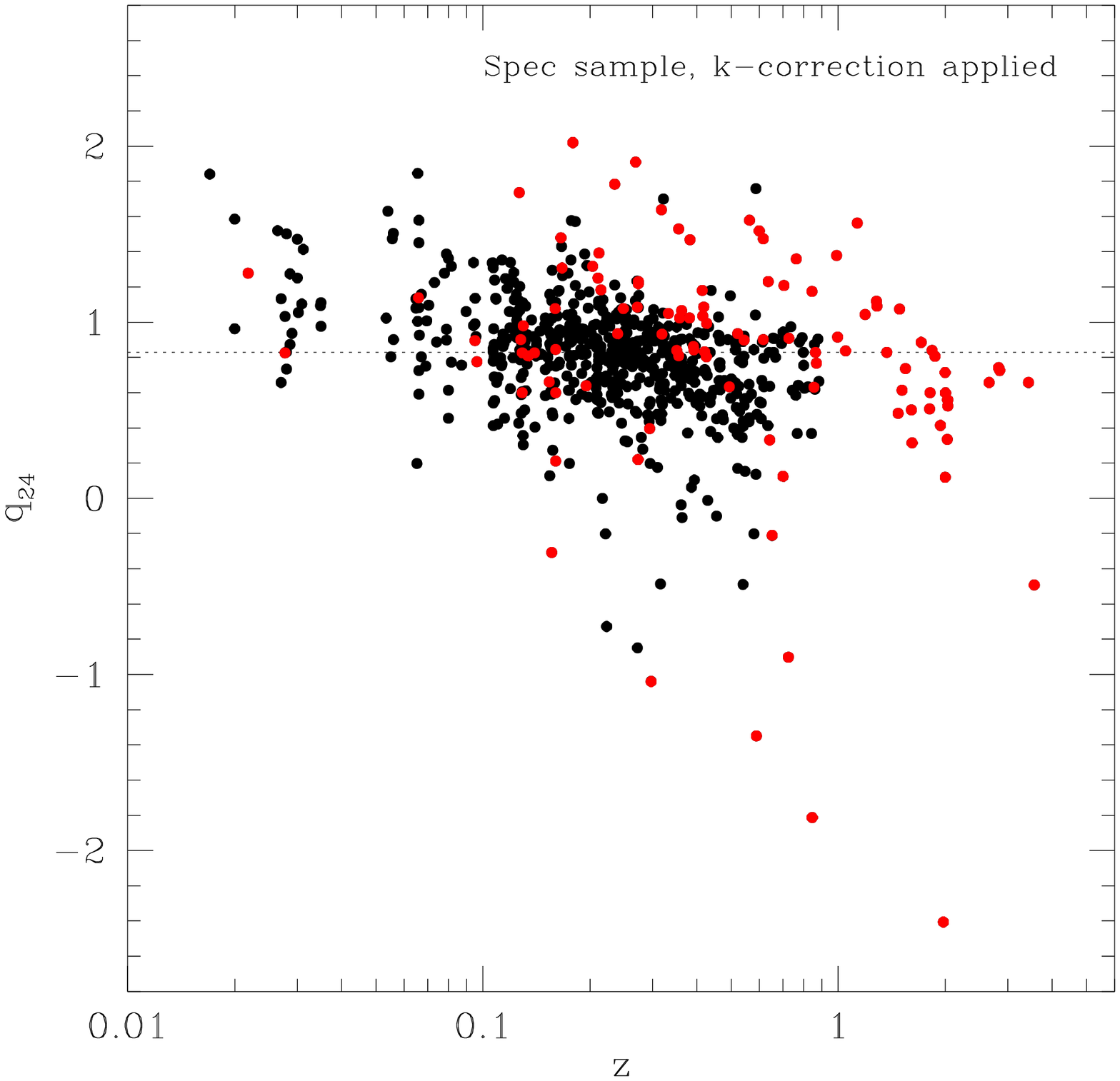}
}
\caption{\label{fig:q24} {\em Left}: Relation between 24~\mum\ and
1.4~GHz raw flux densities, i.e.\ no k-correction applied ({\em open
circles}). The sources with redshifts are shown with {\em filled
black circles} and the optically classified AGNs are identified as
{\em filled red circles}. The {\em dotted line} represents the median
value of q$_{24}$ = 0.89, measured using the uncorrected flux densities. 
{\em Right}: The parameter q$_{24}$, computed using k-corrected flux densities, 
as a function of redshift for star-forming galaxies ({\em filled black 
circles}). The AGNs that we have identified in the WIYN/Hydra, SDSS 
and MMT catalogs are shown as {\em filled red circles}. The {\em dotted 
line} represents the median value of q$_{24}$ = 0.83.}
\end{figure}

For star-forming galaxies, there exists a tight correlation between
far-infrared and radio luminosity. However, as shown by Appleton et
al.\ (2004), this correlation shows much larger scatter when 24~\mum\
flux densities are used. The monochromatic 24~\mum\ infrared-radio
correlation can best be plotted as a function of redshift by computing
the parameter

\begin{equation}
q_{24} = \log(S_{24{\mu}m}/S_{1.4~GHz}),
\end{equation}

\noindent where $S_{24{\mu}m}$ and $S_{1.4~GHz}$ is the flux density
in the 24~\mum\ band and at 1.4~GHz, respectively. The median value of
q$_{24}$ is a measure of the slope of the infrared-radio correlation,
whereas the dispersion is related to the strength of the correlation
for the given sample of galaxies. For our WIYN/Hydra sources, the
1.4~GHz and 24~\mum\ raw flux densities, i.e.\ no k-correction
applied, are plotted against each other in Figure~\ref{fig:q24}
(left). Using these uncorrected flux densities, we calculated the
median value of q$_{24}$ to be 0.89, with a dispersion of 0.30. The
values of q$_{24}$ were also computed using k-corrected flux densities
and are plotted against redshift in Figure~\ref{fig:q24}
(right). Removing the optically classified AGNs, we measured the
median value to be q$_{24}$ = 0.83, with a 1-sigma dispersion of 0.31,
in agreement with Appleton et al.\ (2004). This is not too surprising
as there is some overlap between our sample and the one of Appleton et
al.\ (2004), although the latter is derived mostly from Keck/DEIMOS
redshifts (see Choi et al.\ 2006). From Figure~\ref{fig:q24} (right),
we can see that there is clearly a population of radio-loud AGN
objects which falls off the correlation. These sources were not
classified as optical AGNs. They are possibly buried AGNs, or sources
that have not been corrected properly using the M82 template for the
24~\mum\ luminosity calculation. The former explanation is most
probably correct, as this population is also seen when the raw flux
densities are plotted (Figure~\ref{fig:q24}, left). We also note a
slight tendancy for the radio-quiet AGNs to lie below the average
correlation, i.e.\ higher q$_{24}$ (see Figure~\ref{fig:q24},
left). This effect, however, is less noticeable when the k-correction
is applied (see Figure~\ref{fig:q24}, right). It is not clear how
significant this is, since the k-corrected 24~\mum\ flux densities
depend on the assumed infrared SED, mostly for the high redshift
sources.

\section{Radio and Infrared Luminosity Functions}

The 1.4~GHz and 24~\mum\ luminosity functions (LFs) are computed using
the $1/V_{max}$ method (Schmidt 1968). $V_{max}$ is the maximum volume
within which a galaxy at redshift $z$ could be detected by our survey,
given the flux density limits of $f^{lim}_{1.4~GHz} =$ 0.09~mJy and
$f^{lim}_{24{\mu}m} =$ 0.3~mJy, and the optical limits of R $<$ 23
(WIYN/Hydra) and $i' < 21.3$ (SDSS and MMT/Hectospec). The maximum
comoving volume is given by

\begin{equation}
V_{max} = (\omega/3) d_M^3,
\end{equation}

\noindent where $\omega$ is the solid angle subtended by the FLS 4.4
sq.\ deg.\ area, and $d_M$ is the proper motion distance defined as

\begin{equation}
d_M = \frac{c}{H_0} \int_{z_{min}}^{z_{max}} \frac{dz}{\sqrt{ (1+z)^2 (1+\Omega_M z)-z (2+z) \Omega_{\Lambda} }}.
\end{equation}

\noindent Here, $z_{min}$ and $z_{max}$ refer to the minimum and
maximum redshifts (see, e.g., Carroll, Press \& Turner 1992). In our
survey, $z_{min} = 0$ and $z_{max}$ can be computed from
$d_L(z_{max})=d_L(z)$ [$f_{24{\mu}m}/f^{lim}_{24{\mu}m}$] (IR sample),
or $d_L(z_{max})=d_L(z)$ [$f_{1.4~GHz}/f^{lim}_{1.4~GHz}$] (radio
sample), where $d_L(z)$, the luminosity distance at redshift $z$, is
equal to $(1+z) d_M(z)$.

The density of sources in a given luminosity bin of width $\Delta \log
L$, centered on luminosity $\log L_i$, is simply the sum of the inverse
volumes $1/V_{max}$ of all the sources with luminosities in the
bin. The luminosity function is expressed as

\begin{equation}
\phi(\log L_i) = \frac{1}{\Delta \log L} \sum_{| \log L_j - \log L_i | < \Delta \log L} \frac{1}{V_{max,j}},
\end{equation}

\noindent where the index $i$ labels luminosity bins, the index $j$
labels galaxies, and \\
$1/V_{max,j}$=min($V_{optical,j},V_{24{\mu}m,j}$) (IR sample), or
$1/V_{max,j}$=min($V_{optical,j},V_{1.4~GHz,j}$) (radio sample). Since
Poisson statistics apply, the root-mean-square error is

\begin{equation}
\Delta \phi(\log L_i) = \frac{1}{\Delta \log L} \sqrt{\sum_{| \log L_j - \log L_i | < \Delta \log L} \frac{1}{V^2_{max,j}}}.
\end{equation}

\begin{figure}
\centerline{
\includegraphics[width=240pt,height=240pt,angle=0]{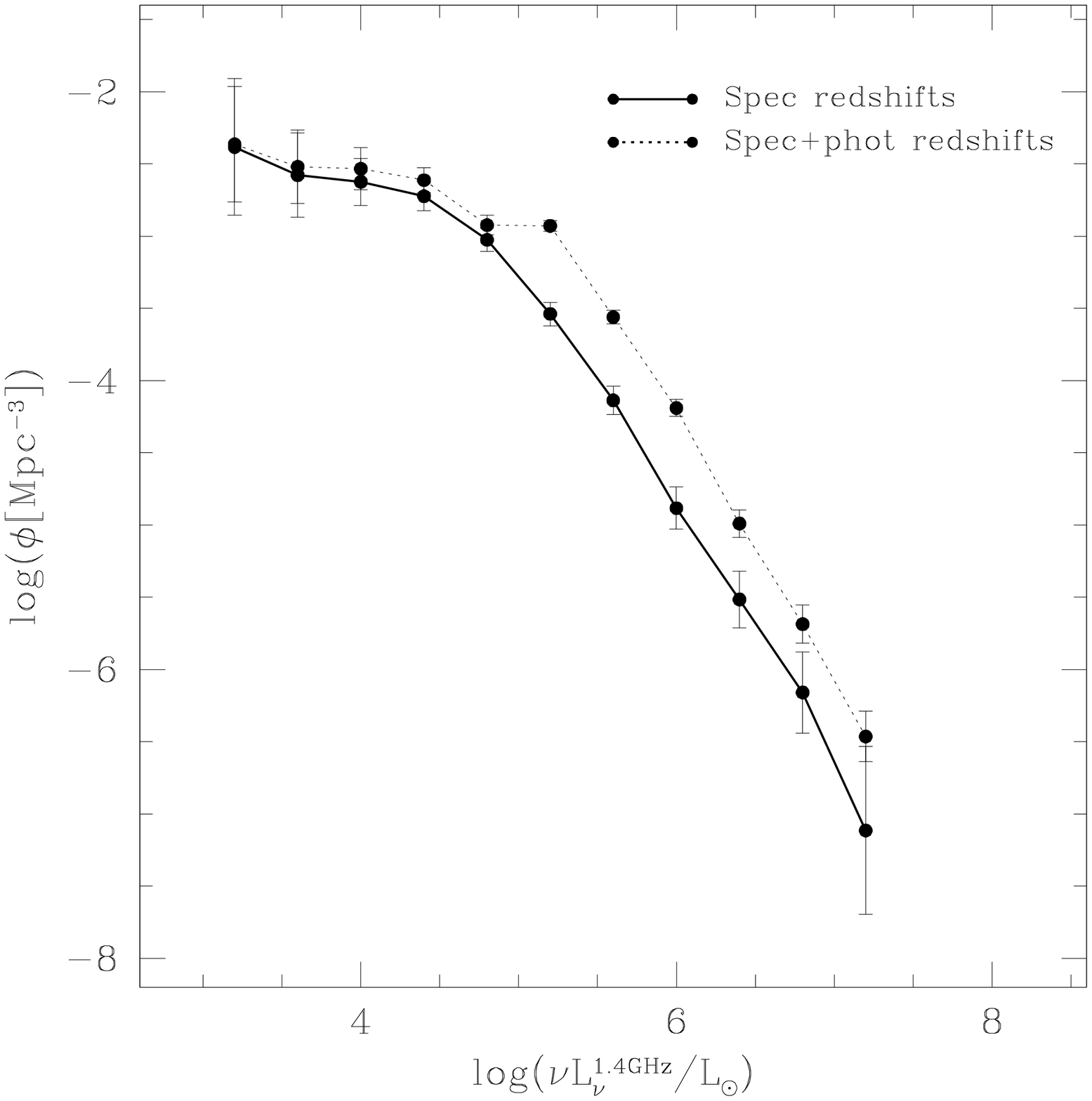}
\includegraphics[width=240pt,height=240pt,angle=0]{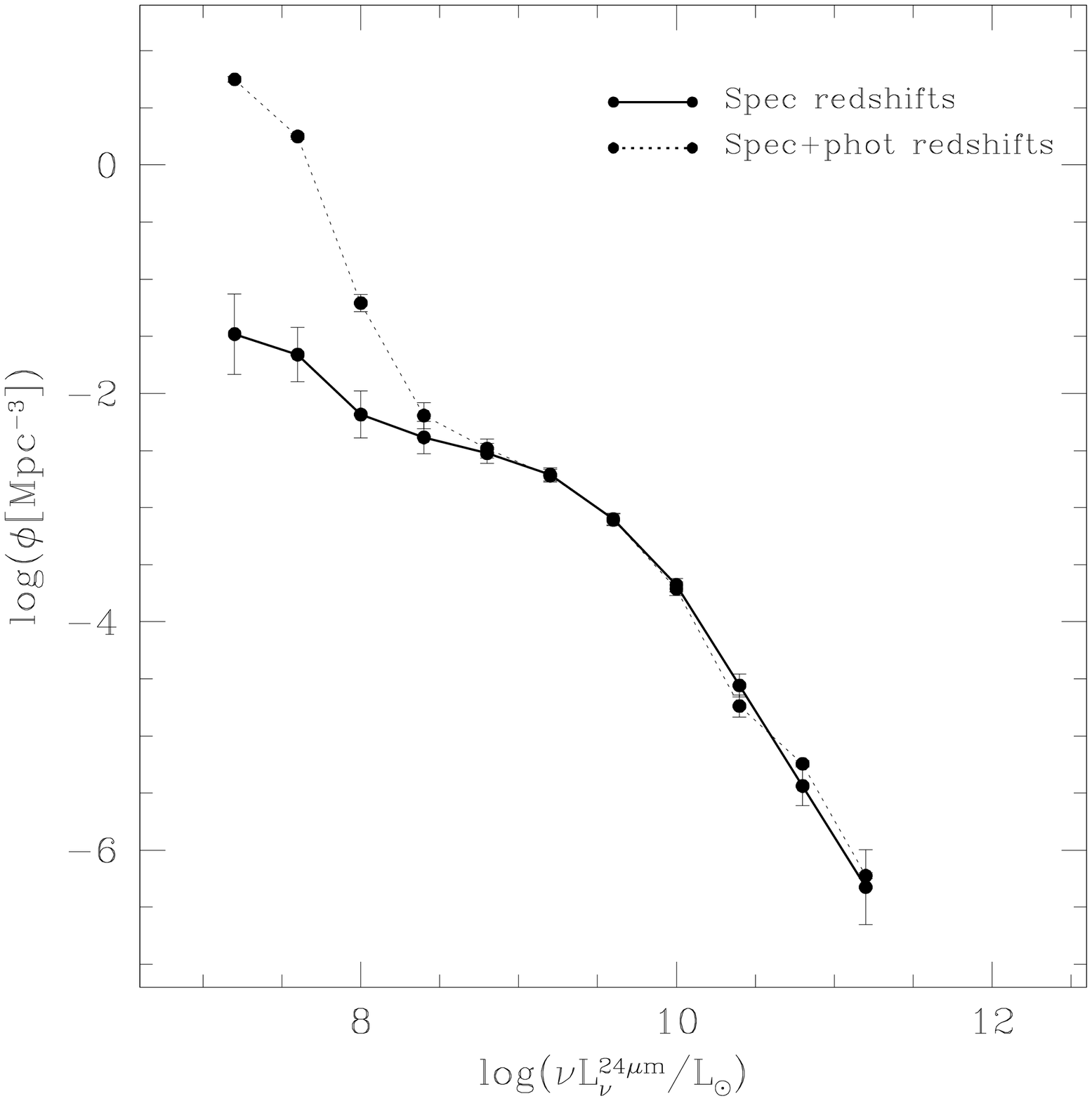}
}
\caption{\label{fig:lf} The 1.4~GHz ({\em left}) and 24~\mum\ ({\em
right}) luminosity function for star-forming galaxies in the FLS. Note
that incompleteness corrections have not been made ({\em solid
lines}). The luminosity functions with added SDSS photometric
redshifts, where the AGN component was removed statistically, 
are shown as the {\em dotted lines}.}
\end{figure}

In Figure~\ref{fig:lf} we show our derived 1.4~GHz and 24~\mum\
luminosity functions for star-forming galaxies in the FLS. A total of
125 and 298 AGNs were removed from the radio and 24~\mum\ all redshift
catalogs, respectively. The luminosity function is plotted as a
function of $\log(\nu L_{\nu}/L_{\odot})$, with bin width defined as
$\Delta \log(\nu L_{\nu}/L_{\odot})=0.4$. The values are listed in
Table~\ref{tbl:lfvla} and \ref{tbl:lf24um}. The luminosity functions
presented in this paper have not been corrected for incompleteness. A
future paper will be dedicated to a more indepth study of the
luminosity functions, and will exploit a new set of spectra of fainter
galaxies (obtained with the Keck telescope) and photometric redshift
estimates for the remaining sources.

Nevertheless, a preliminary estimate of the incompleteness correction
at low redshift is presented here. We queried the SDSS archive for
photometric redshifts in the FLS region\footnote{The SDSS is 95\%
complete at r'=22.2 and i'=21.3 (Abazajian et al.\ 2004),
corresponding to R=21.8 (Fadda et al.\ 2004).}. A total of 58321
sources with SDSS photometric redshifts were retrieved. We matched our
24~\mum\ and 1.4~GHz source lists with the SDSS photometric redshift
catalog and found 8441 and 2348 matches, respectively. As the SDSS
photometric redshift catalog is contaminated by AGNs, we corrected the
luminosity function with added photometric redshifts by applying a
correction factor, derived from the measured fraction of AGNs per
luminosity bin in our spectroscopic sample. The results are shown in
Figure~\ref{fig:lf} as the dotted lines. As we are less complete at
1.4~GHz than at 24~\mum\ (see Figure~\ref{fig:zfracvla} and
\ref{fig:zfrac24um}), the completeness correction is more important in
the radio.

\begin{figure}
\centerline{
\includegraphics[width=240pt,height=240pt,angle=0]{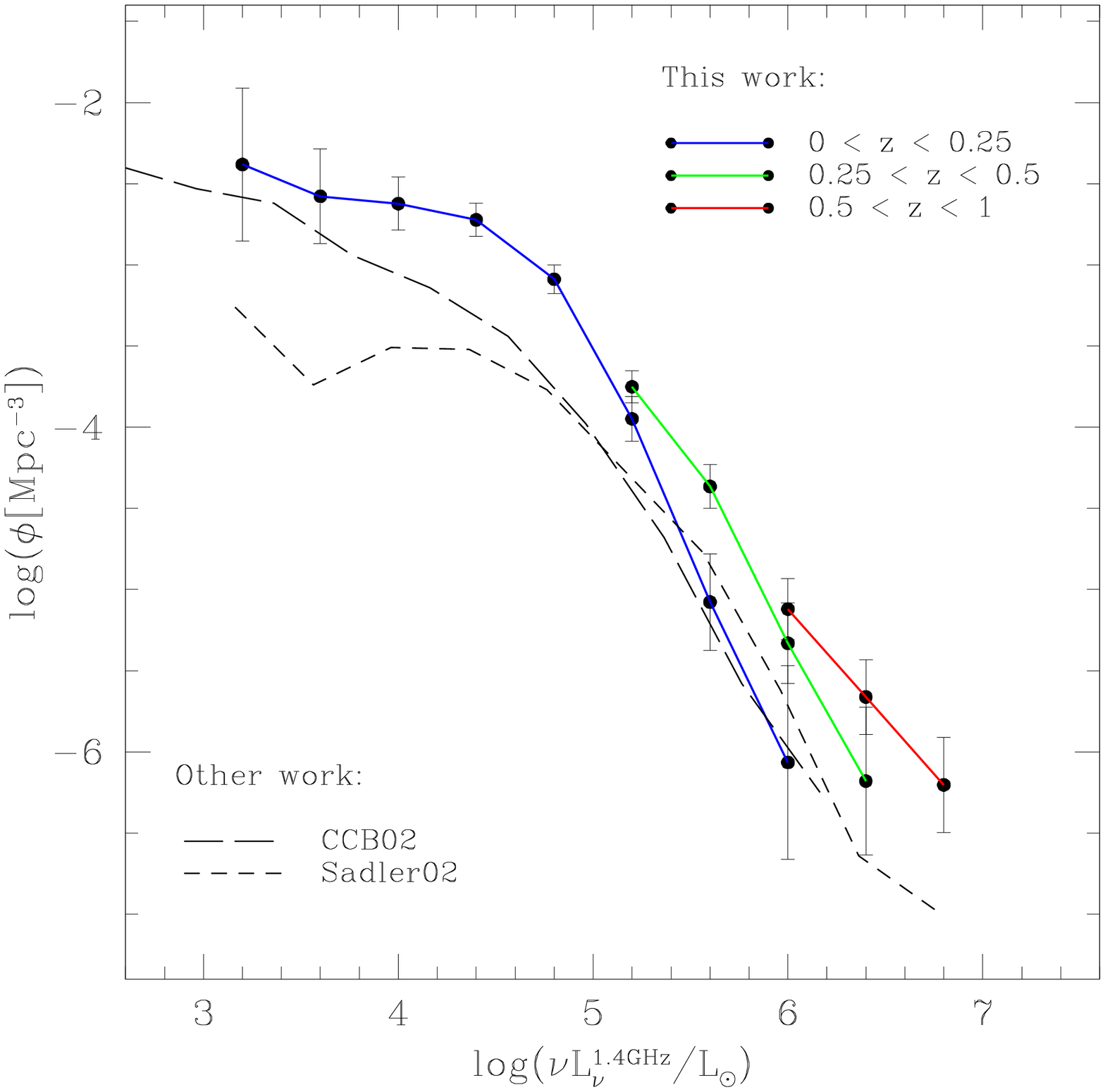}
\includegraphics[width=240pt,height=240pt,angle=0]{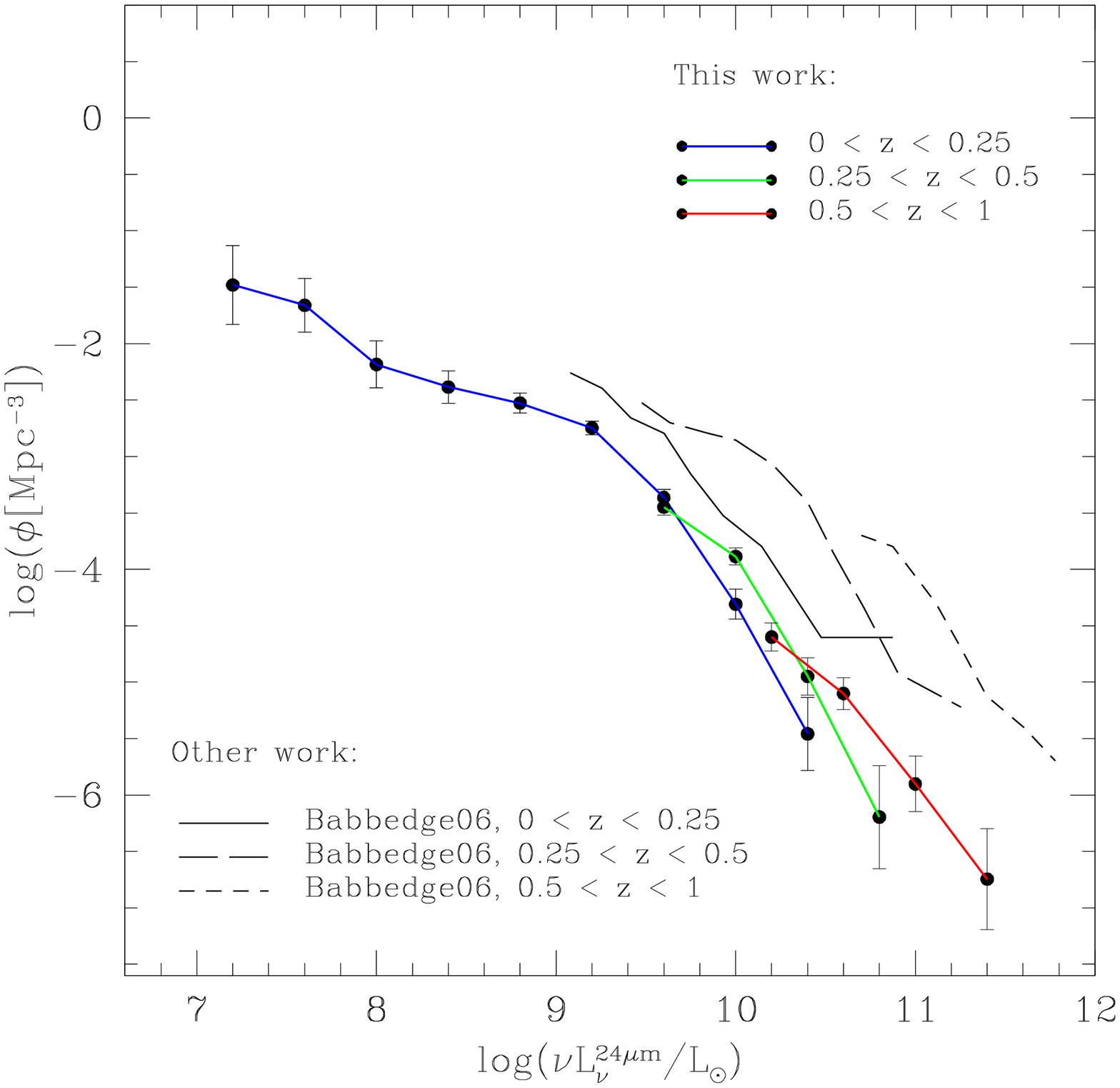}
}
\caption{\label{fig:lfz} The 1.4~GHz ({\em left}) and 24~\mum\ ({\em
right}) luminosity function for star-forming galaxies in the FLS,
split into the redshift bins $z = 0 - 0.25$ ({\em blue line}), $0.25 -
0.5$ ({\em green line}), and $0.5 - 1.0$ ({\em red line}). On the {\em
left}, our 1.4~GHz luminosity function is compared to the 1.4~GHz
local luminosity function of Condon, Cotton \& Broderick (2002) ({\em
long-dash line}) and the luminosity function of Sadler et al.\ (2002) 
({\em short-dash line}). On the {\em right}, our 24~\mum\ luminosity 
function is compared to the 24~\mum\ luminosity function of Babbedge 
et al.\ (2006) for the redshift bins $z = 0 - 0.25$ ({\em solid line}), 
$0.25 - 0.5$ ({\em long-dash line}), and $0.5 - 1.0$ ({\em short-dash 
line}).}
\end{figure}

\begin{figure}
\centerline{
\includegraphics[width=240pt,height=240pt,angle=0]{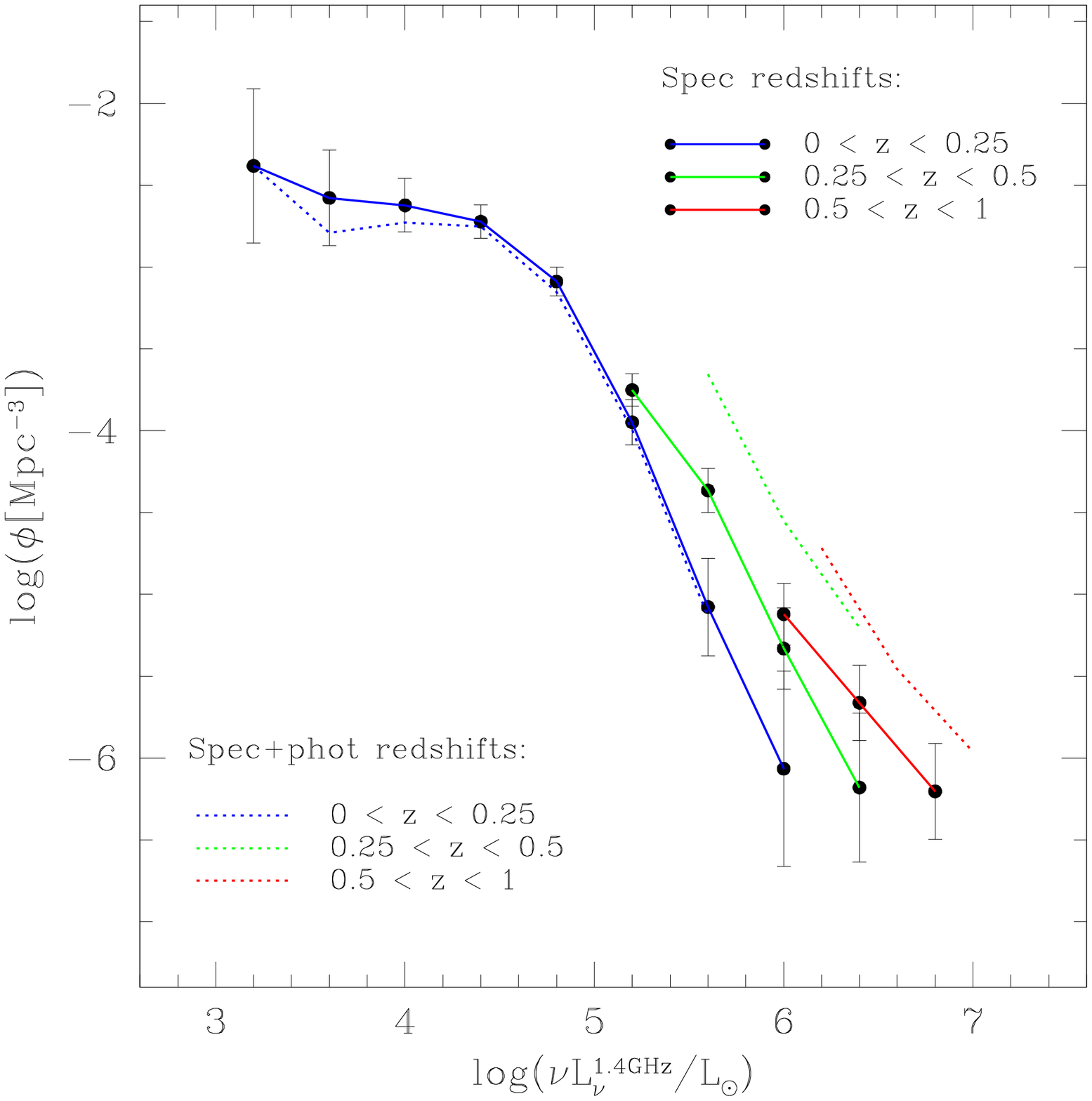}
\includegraphics[width=240pt,height=240pt,angle=0]{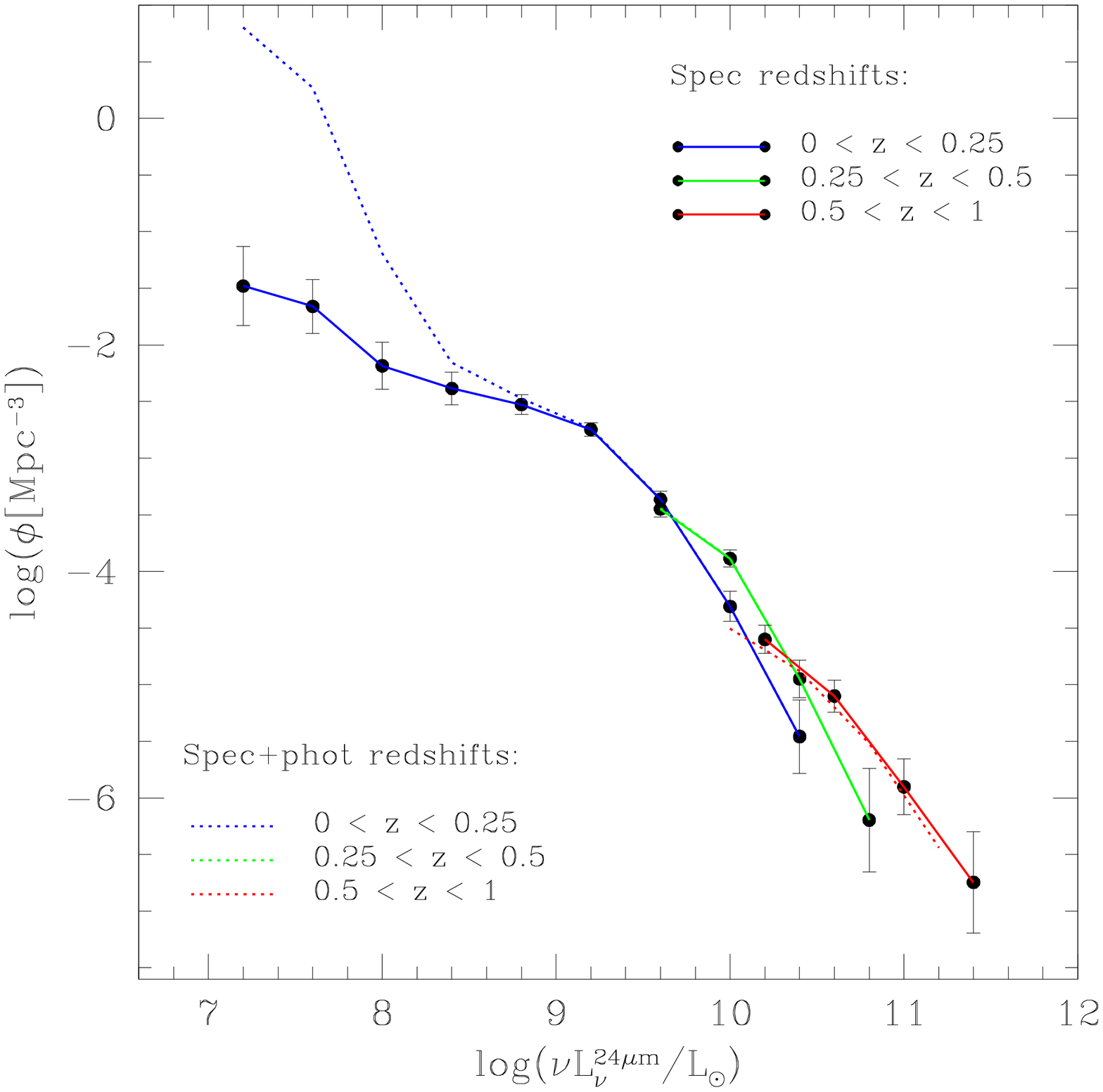}
}
\caption{\label{fig:lfzphot} The 1.4~GHz ({\em left}) and 24~\mum\
({\em right}) luminosity function for star-forming galaxies in the
FLS, split into the redshift bins $z = 0 - 0.25$ ({\em blue line}),
$0.25 - 0.5$ ({\em green line}), and $0.5 - 1.0$ ({\em red line}). The
{\em dotted colored lines} refer to the luminosity function with added
SDSS photometric redshifts, where the AGN component was removed 
statistically, for the same redshift bins.}
\end{figure}

Both luminosity functions are split into the redshift bins $z = 0 -
0.25$, $0.25 - 0.5$, and $0.5 - 1.0$, to investigate evolution in the
24~\mum\ and 1.4~GHz sample (see Figure~\ref{fig:lfz}). We compare our
24~\mum\ luminosity function with the luminosity function of Babbedge
et al.\ (2006), based on photometric redshifts. Our luminosity
functions are lower than those of Babbedge et al.\ (2006), even when
the photometric redshifts are included (Figure~\ref{fig:lfzphot},
dotted lines). This is not surprising, as our luminosity functions are
not corrected for incompleteness. The 1.4~GHz luminosity function
derived for the redshift bin $z = 0 - 0.25$ agrees well with the
1.4~GHz local luminosity function of Condon, Cotton \& Broderick
(2002) and Sadler et al.\ (2002) at the bright end, and is larger at
fainter luminosities. This difference can be explained by the lower
flux density limit of our survey (0.09~mJy compared to
2.5~mJy). Moreover, the greater depth of our radio survey, compared to
previous surveys, allows us to present the first 1.4~GHz luminosity
functions at high redshift (see Figure~\ref{fig:lfz}). Luminosity
evolution over the redshift range $z = 0 - 1$ is clearly detected in
both 1.4~GHz and 24~\mum\ luminosity functions, as shown in
Figure~\ref{fig:lfz}. This is very interesting as it shows that the
infrared and radio galaxy populations were forming stars more actively
in the past, like the ultraviolet and optical populations (e.g., Lilly
et al.\ 1995, 1996; Madau et al.\ 1996), and that therefore this
increase in activity at earlier times appears to be a generic property
of galaxies.

We also make use of the $V/V_{max}$ method (Schmidt 1968) to test the
evolution of the uniformity of the space distribution of sources. The
volume, $V$, is the volume corresponding to the actual redshift at
which the source is observed and, therefore, the ratio $V/V_{max}$ is
a measure of the position of the source within the observable volume,
$V_{max}$. The distribution of sources is uniform if $V/V_{max}$ has a
uniform distribution from 0 to 1, or the mean value of $V/V_{max} =
0.5$. We compute a mean value of $V/V_{max} = 0.34 \pm 0.26$ and $0.41
\pm 0.24$ at 24~\mum\ and 1.4~GHz, respectively. Although the upper
limits are consistent with a luminosity evolution, their low values
are indicative of completeness issues and, therefore, cannot be used
reliably until incompleteness corrections have been applied.

\section{Database of Spectra}

The electronic edition of this publication contains the WIYN/Hydra
redshift catalog and line measurements, i.e.\ Table~\ref{tbl:speccat},
\ref{tbl:lineem} and \ref{tbl:lineext}. The reduced spectra can be
obtained from the {\em Spitzer} FLS website\footnote{
http://ssc.spitzer.caltech.edu/fls/extragal/.}.

\section{Summary}

We presented an extensive redshift survey of 1.4~GHz and 24~\mum\
sources in the FLS main and verification fields. We observed 772
sources, and measured redshifts for 382 radio and 412 infrared
sources, adding up to 498 redshifts. In the deeper FLS verification
region, our survey is 100\% complete for 24~\mum\ flux densities $>$
0.2~mJy. The success rate of redshift identification of 24~\mum\
sources is $\sim$ 80\%. For the radio sources, our redshift success
rate is slightly lower ($\sim$ 60\%). The WIYN/Hydra redshift catalog
and line measurements are available with the online version of this
publication. The reduced spectra can be obtained from the {\em
Spitzer} FLS website.

Our spectroscopic and emission-line survey covered mainly redshifts in
the range $0 < z < 0.4$, with a distribution peaking at $z \sim
0.2$. Most of our WIYN/Hydra sources are galaxies, with eight broad
emission line objects, one of which is a quasar at $z = 3.6$. Using
the diagnostic line ratios, to separate starburst from AGNs, we found
an additional 28 sources with AGN-like emission properties.

The Balmer decrement was used to correct emission line fluxes for
extinction. We calculated the SFRs for the subset of our sample with
measured H$\alpha$ fluxes. We found that star formation rates computed
from the H$\alpha$ line flux measurements ranged from 0.2-200
M$_\odot$ yr$^{-1}$, and correlated directly with extinction. The star
formation rates derived from the radio luminosities were found to be
on average consistent with those estimated using the H$\alpha$ line
fluxes. Moreover, they did not correlate with optical extinction. At
low redshift ($z < 0.4$), our survey detected sources with a wide
range of star-forming activity. However, at higher redshifts, we only
detected sources with a SFR $> 40$ M$_\odot$ yr$^{-1}$.

We complemented our redshift database with the redshifts produced by
the SDSS and MMT surveys, yielding a total of 830 redshifts for the
1.4~GHz sources and 1753 for the 24~\mum\ sources. The 1.4~GHz and
24~\mum\ luminosity of sources in the FLS was computed for all
galaxies with a redshift. The majority of galaxies in the redshift
range $z = 0 - 0.4$ were found to have low to medium radio luminosity,
typical of star-forming galaxies, whereas almost all the higher
redshift galaxies had higher radio luminosities, comparable to
powerful luminous infrared galaxies and radio-quiet quasars. The
infrared-radio correlation was derived and a median value of $q_{24}$
= 0.83 was measured, with a 1-sigma dispersion of 0.31.

The 24~\mum\ and 1.4~GHz luminosity functions -- uncorrected for
incompleteness -- were derived from this spectroscopic sample. By
splitting our sample into the redshift bins $z = 0 - 0.25$, $0.25 -
0.5$, and $0.5 - 1.0$, we found evidence for luminosity evolution in
both the 1.4~GHz and 24~\mum\ luminosity functions.

\acknowledgements

The authors are most grateful to F.\ Valdes for his assistance with
using the Hydra reduction package. This work is based in part on
observations made with the {\em Spitzer Space Telescope}, which is
operated by the Jet Propulsion Laboratory, California Institute of
Technology, under NASA contract 1407. We wish to thank the National
Optical Astronomy Observatories for generous allocation of telescope
time at the WIYN telescope at the Kitt Peak National Observatory. The
WIYN observatory is a joint facility of the University of
Wisconsin-Madison, Indiana University, Yale University, and the
National Optical Astronomy Observatories. We thank an anonymous
referee for helpful suggestions on improving the manuscript. M.I.\ was
supported by the grant No.\ R01-2005-000-10610-0 from the Basic
Research Program of the Korea Science \& Engineering Foundation.

\clearpage

\begin{deluxetable}{cccc}
\tablecaption{The 15 Fields Observed with WIYN/Hydra.\label{tbl:fields}}
\tablewidth{0pt}
\tablecolumns{4}
\tablehead{ \colhead{Field}  &\colhead{$\alpha$ (J2000.0)}  &\colhead{$\delta$ (J2000.0)}  &\colhead{Observation Date (UTC)} }
\startdata
FLS1a02  &17:17:00.000  &+59:46:00.00   &2002 Aug 14\\
FLS1b02  &17:16:52.056  &+59:45:00.00   &2002 Aug 14\\
FLS1c02  &17:21:07.881  &+59:31:00.00   &2002 Aug 14\\
FLS2a02  &17:21:07.881  &+59:31:00.00   &2002 Aug 15\\
FLS2b02  &17:12:52.119  &+59:29:00.00   &2002 Aug 15\\
FLS2c02  &17:16:52.060  &+59:45:00.00   &2002 Aug 15\\
FLS1a03  &17:11:35.376  &+60:04:59.91   &2003 Jun 30\\
FLS1b03  &17:13:45.307  &+60:12:01.32   &2003 Jun 30\\
FLS1c03  &17:11:16.641  &+59:39:38.39   &2003 Jun 30\\
FLS2a03  &17:11:18.700  &+59:33:15.20   &2003 Jul 01\\
FLS2b03  &17:17:08.560  &+59:05:48.00   &2003 Jul 01\\
FLS2c03  &17:17:31.600  &+60:01:04.00   &2003 Jul 01\\
FLS2d03  &17:24:48.630  &+59:32:16.00   &2003 Jul 01\\
FLS1a05  &17:25:10.218  &+58:48:15.00   &2005 Jul 17\\
FLS1b05  &17:25:19.624  &+59:59:53.30   &2005 Jul 17\\
\enddata
\end{deluxetable}

\clearpage

\begin{deluxetable}{lll}
\tablecaption{Format of the WIYN/Hydra Spectroscopic Catalog of 1.4~GHz
and 24~Micron Sources in the {\em Spitzer} FLS \label{tbl:speccat}}
\tablewidth{0pt}
\tablecolumns{3}
\tablehead{ \colhead{Column}  &\colhead{Format}  &\colhead{Description} }
\startdata
   1:3     &i3    & ID \\
   5:11    &a7    & Field (as defined in Table~\ref{tbl:fields}) \\
  13:18    &a6    & APID \\
  20:39    &a20   & WIYN/Hydra Target RA,Dec (J2000) \\
  41:49    &f7.4  & WIYN/Hydra Redshift \\
  49:54    &f6.4  & WIYN/Hydra Redshift Error \\
  56:56    &i1    & WIYN/Hydra Redshift Quality \\
           &      & [0:unknown, 1:tentative, 2:good] \\
  58:60    &a3    & WIYN/Hydra Type of Spectrum \\
           &      & [sta:star, gal:galaxy, bel:broad emission line] \\
  62:67    &a6    & WIYN/Hydra Features \\
           &      & [b:break, e:emission, a:absorption] \\
  69:74    &f6.2  & Optical Counterpart R-Band Magnitude \\
  76:95    &a20   & Radio 1.4~GHz/20~cm Counterpart RA,Dec (J2000) \\
  97:103    &f7.3  & Radio Counterpart 1.4~GHz/20~cm Flux Density [mJy] \\
  105:110  &f6.3  & Radio Counterpart 1.4~GHz/20~cm Flux Density Error [mJy] \\
  112:116  &f5.2  & WIYN/Radio Match Separation [arcsec] \\
  118:137  &a20   & FIR 24~\mum\ Counterpart RA,Dec (J2000) \\
  139:143  &f5.2  & FIR Counterpart 24~\mum\ Flux Density [mJy] \\
  145:149  &f5.2  & FIR Counterpart 24~\mum\ Flux Density Error [mJy] \\
  151:155  &f5.2  & WIYN/FIR Match Separation [arcsec] \\
  157:157  &a1    & FIR Source Type \\
           &      & [E:extended, P:point source, -:no match] \cr
  159:165  &f7.4  & SDSS Redshift \cr
  167:173  &f7.4  & MMT Redshift \cr
\enddata
\end{deluxetable}

\begin{deluxetable}{lll}
\tablecaption{Format of the Uncorrected Line Strengths and Equivalent Widths 
Table \label{tbl:lineem}}
\tablewidth{0pt}
\tablecolumns{3}
\tablehead{ \colhead{Column}  &\colhead{Format}  &\colhead{Description} }
\startdata
   1:3     &i3    & ID \\
   5:11    &a7    & Field (as defined in Table~\ref{tbl:fields}) \\
  13:18    &a6    & APID \\
  20:39    &a20   & WIYN/Hydra Target RA,Dec (J2000) \\
  41:46    &f6.1  & CIII $\lambda$1909 Flux [1e-17 erg/cm$^2$/s]\\
  48:53    &f6.1  & CIII $\lambda$1909 EW [\AA]\\
  55:60    &f6.1  & MgII $\lambda$2798 Flux [1e-17 erg/cm$^2$/s]\\
  62:67    &f6.1  & MgII $\lambda$2798 EW [\AA]\\
  69:74    &f6.1  & H$\beta$ $\lambda$4861 Flux [1e-17 erg/cm$^2$/s]\\
  76:81    &f6.1  & H$\beta$ $\lambda$4861 EW [\AA]\\
  83:88    &f6.1  & [OIII] $\lambda$5007 Flux [1e-17 erg/cm$^2$/s]\\
  90:95    &f6.1  & [OIII] $\lambda$5007 EW [\AA]\\
  97:102   &f6.1  & H$\alpha$ $\lambda$6563 Flux [1e-17 erg/cm$^2$/s]\\
  104:109  &f6.1  & H$\alpha$ $\lambda$6563 EW [\AA]\\
  111:116  &f6.1  & [NII] $\lambda$6583 Flux [1e-17 erg/cm$^2$/s]\\
  118:123  &f6.1  & [NII] $\lambda$6583 EW [\AA]\\
  125:130  &f6.1  & [SII] $\lambda$6716 Flux [1e-17 erg/cm$^2$/s]\\
  132:137  &f6.1  & [SII] $\lambda$6716 EW [\AA]\\
  139:144  &f6.1  & [SII] $\lambda$6731 Flux [1e-17 erg/cm$^2$/s]\\
  146:151  &f6.1  & [SII] $\lambda$6731 EW [\AA]\\
\enddata
\end{deluxetable}

\clearpage 

\begin{deluxetable}{lll}
\tablecaption{Format of the Dereddened Line Ratios and Extinction 
Table \label{tbl:lineext}}
\tablewidth{0pt}
\tablecolumns{3}
\tablehead{ \colhead{Column}  &\colhead{Format}  &\colhead{Description} }
\startdata
   1:3     &i3    & ID \\
   5:11    &a7    & Field (as defined in Table~\ref{tbl:fields}) \\
  13:18    &a6    & APID \\
  20:39    &a20   & WIYN/Hydra Target RA,Dec (J2000) \\
  42:47    &f7.2  & Balmer Decrement ($F_o^{H\alpha}/F_o^{H\beta}$) \\
  49:55    &f7.2  & Color Excess E(B-V) [mag] \\
  57:63    &f7.2  & Extinction A(V) [mag] \\
  65:71    &f7.2  & Dereddened H$\alpha$ Luminosity log([erg/s]) \\
  73:79    &f7.2  & 1.4~GHz Luminosity log([W/Hz]) \\
  81:87    &f7.2  & 24~\mum\ Luminosity log([W/Hz]) \\
  89:95    &f7.2  & SFR$_{H\alpha}$ [M$_\odot$ yr$^{-1}$] \\
  97:103   &f7.2  & SFR$_{1.4~GHz}$ [M$_\odot$ yr$^{-1}$] \\
  105:111  &f7.2  & log([OIII] $\lambda$5007/H$\beta$ $\lambda$4861) \\
  113:119  &f7.2  & log([NII] $\lambda$6583/H$\alpha$ $\lambda$6563) \\
  121:127  &f7.2  & log([SII] $\lambda$6716+6731/H$\alpha$ $\lambda$6563) \\
\enddata
\end{deluxetable}

\clearpage

\begin{deluxetable}{ccc}
\tablecaption{The 1.4~GHz LF in the FLS from the 1/$V_{max}$ Analysis \label{tbl:lfvla}}
\tablewidth{0pt}
\tablecolumns{3}
\tablehead{ \colhead{$\log(\nu L_{\nu}^{1.4~GHz}/L_{\odot})$}  &\colhead{$\log(\phi [$Mpc$^{-3}])$}  &\colhead{1$\sigma$ error} }
\startdata
 3.2  &-2.38  &0.47\\
 3.6  &-2.58  &0.29\\
 4.0  &-2.62  &0.16\\
 4.4  &-2.72  &0.10\\
 4.8  &-3.02  &0.08\\
 5.2  &-3.54  &0.08\\
 5.6  &-4.14  &0.10\\
 6.0  &-4.88  &0.15\\
 6.4  &-5.52  &0.20\\
 6.8  &-6.16  &0.28\\
 7.2  &-7.12  &0.58\\
\enddata
\end{deluxetable}

\clearpage

\begin{deluxetable}{ccc}
\tablecaption{The 24~\mum\ LF in the FLS from the 1/$V_{max}$ Analysis \label{tbl:lf24um}}
\tablewidth{0pt}
\tablecolumns{3}
\tablehead{ \colhead{$\log(\nu L_{\nu}^{24\mu m}/L_{\odot})$}  &\colhead{$\log(\phi [$Mpc$^{-3}])$}  &\colhead{1$\sigma$ error} }
\startdata
 7.2  &-1.48  &0.35\\
 7.6  &-1.66  &0.24\\
 8.0  &-2.18  &0.21\\
 8.4  &-2.39  &0.14\\
 8.8  &-2.53  &0.09\\
 9.2  &-2.71  &0.06\\
 9.6  &-3.10  &0.05\\
10.0  &-3.68  &0.06\\
10.4  &-4.56  &0.10\\
10.8  &-5.44  &0.17\\
11.2  &-6.32  &0.33\\
\enddata
\end{deluxetable}

\clearpage

\clearpage

\clearpage

\clearpage

\clearpage

\clearpage

\clearpage

\clearpage

\clearpage

\clearpage

\clearpage

\clearpage

\clearpage

\clearpage

\clearpage

\clearpage

\clearpage

\end{document}